\documentclass[journal]{IEEEtran}
\usepackage[T1]{fontenc}
\usepackage{yfonts} 
\usepackage{amsmath}
\usepackage{amssymb}
\DeclareMathOperator{\sign}{sign}

\usepackage{array}
\newcolumntype{M}[1]{>{$\displaystyle\qquad}p{#1}<{$}}
\newcolumntype{C}[1]{>{\centering\let\newline\\\arraybackslash\hspace{0pt}}m{#1}}
\newcolumntype{L}[1]{>{\raggedright\let\newline\\\arraybackslash\hspace{0pt}}m{#1}}

\usepackage{mathtools} 
\usepackage{graphicx}
\usepackage{multirow}
\usepackage{float}
\usepackage[subrefformat=parens,labelformat=parens,caption=false,font=footnotesize]{subfig}
\usepackage[comma,numbers,sort&compress]{natbib}
\usepackage{siunitx}

\usepackage{t1enc}
\usepackage{times}
\usepackage{tikz}
\usetikzlibrary{arrows,arrows.meta,shapes.geometric,positioning,calc,chains,backgrounds}
\tikzset{>={Latex[width=1.7mm,length=1.7mm]}}
\usepackage[siunitx]{circuitikz}

\usepackage{pgfplots}            
\pgfplotsset{compat=1.17}

\usetikzlibrary{external} 
\newif\ifusetikzpdf

\usetikzpdftrue 

\newcommand{\de}[0]{\si{d}}
\newcommand{\ju}{\mathrm{j}\mkern1mu}

\begin{document}

\newcommand\mytodo[1]{\textcolor{red}{#1}}

\raggedbottom
%
%
%
\title{
    Nonlinear Noise Mechanisms in Active Devices: Additive Amplitude Noise and
    Phase Noise
} 
\author{
        Meysam Bahmanian\\
        System and Circuit Technology Group, Heinz Nixdorf Institute,
Paderborn University, 33102 Paderborn, Germany (e-mail:
meysam.bahmanian@uni-paderborn.de)

}
%

%
%
\maketitle
%
%
%
\begin{abstract}
    In this report, we lay the foundation for amplitude noise and phase noise analysis in
    nonlinear devices. We build a theoretical framework that helps us to analyze extremely
    difficult problems which include both nonlinearity and noise of semiconductors. Using
    our proposed framework, we analytically calculate the amplitude noise and phase noise
    of nonlinear RC circuits in the presence of nonlinear capacitance and conductance. 
    As a more practical example, we analyze the amplitude noise and phase noise of a
    common-emitter stage bipolar transistor. 
\end{abstract}
%
%
\IEEEpeerreviewmaketitle

\section{Introduction}\label{sec:int}
Many advanced technological achievements are closely related to our ability to precisely
measure time. The application range is wide: from our scientific experiments in a large
scale aiming for observing the universe \cite{Cliche2006} to experiments that look into
intermolecular forces \cite{Nuhn2002}.
In fundamental research facilities, a sample is exposed to a short pulse of high
energy particles, and the particles generated by this interaction are monitored via
detectors. The overall accuracy of the experiment depends upon the precision of timing the
high energy particles that hit the sample. In engineering field specifically, the 
effective number of bits (ENOB) of high speed analog-to-digital converters (ADC) and
digital-to-analog converters (DAC) has entered a regime that is limited by the jitter of
the clock signal
\cite{Khilo2012}. In heterodyne receivers, the phase noise of clock oscillators (which is
timing jitter normalized to the carrier period) might get frequency-mixed with a strong
interference and entirely mask the desired signal. All these RF systems and subsystems
point to the importance of timing precision. \cite{Scheer1990}

It is well known that both phase noise and amplitude noise are closely related to
nonlinear mechanisms in active devices. One of the first notable works in phase noise
analysis was done by \citet{Walls1997} (also reported in \cite{Walls2001}) using
perturbation theory, which assumes the noise processes are slow and perturb the amplitude
and phase of the nonlinear device. This consequently leads to generation of amplitude and
phase noise. This approach was later generalized by
\citet{Tomlin2000} to a nonlinear system with multiple noise sources.
\citet{Ferre-Pikal2008} suggested a more practical approach for amplitude and phase noise
estimation. They modeled the gain and phase of a common-emitter stage amplifier as a
function of the bias current. Since the flicker noise of bipolar transistor is
directly proportional to the base current, the sensitivity of the amplitude and phase of
the response to the base current can be used to estimate the amplitude and phase noise of
the transistor. Although this approach provided an accurate estimation, but cannot
give an insight about the dynamics of amplitude and phase noise generation in the device,
which is necessary for optimizing the design for low phase noise. \citet{Boudot2012} tried
to relate the phase noise of the amplifier to simple memoryless 2nd order nonlinearity,
but in the absence of memory elements (like capacitor), this only leads to amplitude
noise. Although this approach can lead to optimization techniques for lower phase
noise, but, as we will show in this report, is fundamentally unable to show the origin of
phase noise in active devices. \citet{Ferre-Pikal2004} used an active feedback technique
that senses the noise at the collector of the transistor and feeds a fraction of the the
sensed current back
to the base to reduce the phase noise. The results show an improvement in the
phase noise by approximately \si{10~dB}, however, the mechanism that leads to phase noise
is not analyzed. Unfortunaltely, due to complexity of this problem, phase
noise in active devices,
which deals with both nonlinearity and random processes, there are not
many analytical works in the literature.

In this paper, we demonstrate that amplitude and phase noise are closely related to
second-order nonlinearities of the devices. We propose a systematic approach for
amplitude/phase noise analysis to incorporate the nonlinear coefficients responsible for
amplitude/phase noise generation. We then analyze the response of several simple nonlinear
elements in
presence of noise. We adapt this approach to nonlinear circuits that do not have a
closed-form response, but rather, only a nonlinear governing differential equation of the
circuit is available which does not have a closed-form response. It is shown that, with
some reasonable approximations, such nonlinear equations can be reduced to a set of
coupled homogeneous differential equations which have an analytic solution.
Although the analysis is complex, the results are simple and easy to understand.
This approach can then be adapted for amplitude/phase noise analysis of a
wide range of nonlinear circuits.

This paper is organized as follows: in section~\ref{sec:basics_apn} we provide basic
mathematical model for phase noise and amplitude noise. In section~\ref{sec:nl_dev} we
extend the linear model of an active device to include its additive phase noise and
amplitude noise. We then calculate the amplitude/phase noise of simple nonlinear elements.
In section~\ref{sec:nl_rc} we analyze three nonlinear RC-circuits that are closely
related to practical applications and find the expressions for their amplitude and phase
noise. As a practical example, in Section~\ref{sec:bipolar}, we apply the results from our
analysis to a common-emitter stage bipolar transistor and the analytical results are
compared with the simulation results. 

\section{Phase Noise and Amplitude Noise} 
\label{sec:basics_apn} 
In this section we look at the mathematical model of phase noise and amplitude noise. This
mathematical model is the basis of our analysis of additive amplitude and phase noise in
nonlinear devices.

Phase noise (or PM noise) is defined as the random variations of the phase of a sinusoidal
signal $x(t)$ as
\begin{flalign}\label{eq:basics_apn_pn_xt}
    x(t) = X_1 \cos\big( \omega_0 t+\phi_n(t) \big),
\end{flalign}
where $X_1$ is the amplitude, $\omega_0$ is the angular frequency, $\phi_0$ is the offset
phase and $\phi_n(t)$ is phase noise. 

%
Like any other random process, it is meaningful to talk about the statistical properties
of phase noise rather than its instantaneous value. Assuming phase noise is ergodic 
(that its temporal averages are equal to ensemble averages), the autocorrelation function of the
phase noise can be written as
\begin{flalign}\label{eq:basics_apn_pn_rpt}
    R_{\phi_n}(\tau) = \textrm{E}[\phi_n(t)\phi_n(t+\tau)],
\end{flalign}
where $\textrm{E}[.]$ denotes the expectation value. The power spectral density of the
phase noise according to Wiener–Khinchin theorem is the Fourier transform of its
autocorrelation function
\begin{flalign}\label{eq:basics_apn_pn_spf}
    S_{\phi_n}(f) = \int_{-\infty}^{+\infty}\de \tau R_{\phi_n}(\tau)e^{-\ju2\pi f\tau}.
\end{flalign}
The PSD of phase noise has units of \si{Rad\squared/Hz}. In the literature, usually what
is referred to as phase noise is the power spectral density of the phase noise. 

Phase noise can also be described as a noise term modulated by a term that has
\si{90}-degree phase difference relative to the carrier, the so-called quadrature carrier.
This can be shown by expansion of the sinusoidal term in \eqref{eq:basics_apn_pn_xt} and
assuming the variance of phase noise is small, $|\phi_n(t)| \ll \pi/2$, as
\begin{flalign}\label{eq:basics_apn_pn_xt_expand}
    x(t) \approx 
    X_1\cos\big(\omega_0 t\big)-
    X_1 \phi_n(t) \sin\left( \omega_0 t \right),
\end{flalign}
where we approximated $\cos(\phi_n(t)) \approx 1$ and $\sin(\phi_n(t)) \approx \phi_n(t)$.
Therefore, the carrier quadrature component carries the information about the phase noise.

%
Amplitude noise (or AM noise) of a signal is random variations of its amplitude with respect to time.
In a similar fashion the phase noise was
included in a noiseless tone, the amplitude noise can also be added to a sinusoidal signal as
\begin{flalign}\label{eq:basics_apn_an}
    x(t) = X_1\big(1+a_n(t)\big) \cos(\omega_0 t),
\end{flalign}
where $a_n(t)$ is the amplitude noise. In contrast to phase noise that manifests itself as
a noise term modulated by the quadrature carrier, amplitude noise is
modulated by the carrier itself. Similar to what has been done for the phase noise,
the autocorrelation function and the PSD of the amplitude noise can be derived
\begin{flalign}\label{eq:basics_apn_rat}
    R_{a_n}(\tau) = \textrm{E}[a_n(t)a_n(t+\tau)],
\end{flalign}
and 
\begin{flalign}\label{eq:basics_apn_saf}
    S_{a_n}(f) = \int_{-\infty}^{+\infty}\de \tau R_{a_n}(\tau)e^{-\ju2\pi f\tau}.
\end{flalign}

Now let's consider a general signal that has both in phase and quadrature noise components
\begin{flalign}\label{eq:basics_apn_xt_ninq}
    x(t) = X_1 \cos(\omega_0 t) + n_{\textbf{I}}(t)\cos(\omega_0 t) + n_{\textbf{Q}}(t)\sin(\omega_0 t)  ,
\end{flalign}
where $n_{\textbf{I}}(t)$ and $n_{\textbf{Q}}(t)$ represent the in-phase- and quadrature-modulated noise terms,
respectively. We want to determine the amplitude noise and phase noise of $x(t)$. Assuming
$|n_{\textbf{I}}(t)| \ll X_1$, \eqref{eq:basics_apn_xt_ninq} can be rewritten as 
\begin{flalign}\label{eq:basics_apn_xt_ninq_aprx}
    \nonumber
    x(t) 
    &\approx X_1 \left( 1+\frac{n_{\textbf{I}}(t)}{X_1} \right)
    \left( \cos(\omega_0 t) + \frac{n_{\textbf{Q}}(t)}{X_1} \sin(\omega_0 t) \right) \\ 
    &\approx X_1 \left( 1+\frac{n_{\textbf{I}}(t)}{X_1} \right)
    \cos\left( \omega_0 t - \frac{n_{\textbf{Q}}(t)}{X_1} \right).
\end{flalign}
Therefore, the ratio of the in-phase noise term to carrier amplitude defines the amplitude
noise and the ratio of the quadrature noise term to carrier amplitude defines the phase
noise, or mathematically
\begin{flalign}\label{eq:basics_apn_xt_apn}
    x(t) = X_1 \big( 1+a_n(t) \big)\cos\big( \omega_0 t +\phi_n(t) \big)
\end{flalign}
where
\begin{flalign}\label{eq:basics_apn_anpn}
    a_n(t) = \frac{n_{\textbf{I}}(t)}{X_1} 
    \quad \text{and} \quad 
    \phi_n(t) = -\frac{n_{\textbf{Q}}(t)}{X_1}.
\end{flalign}

\section{Noise in nonlinear devices} \label{sec:nl_dev}
What is the source of amplitude or phase noise in active devices? We already know that the
additive noise at the output of the active devices has both in-phase and quadrature
carrier components and can be written as 
\begin{flalign}\label{eq:nl_dev_lin_noise}
    \nonumber
    x(t) &= X_{1} \cos(\omega_0 t) + n(t) \\
    &= X_1 \cos(\omega_0 t) + n_{\textbf{I}}(t)\cos(\omega_0 t) + n_{\textbf{Q}}(t)\sin(\omega_0 t)  ,
\end{flalign}
where $n(t)$ is the output noise and $n_{\textbf{I}}(t)$ \& $n_{\textbf{Q}}(t)$ are its carrier in-phase
and quadrature components. These
components have similar statistical properties and therefore equal PSDs.
\begin{flalign}\label{eq:nl_dev_lin_iq}
    S_{n_{\textbf{I}}}(\omega) = S_{n_{\textbf{Q}}}(\omega) = \frac12 S_{n}(\omega).
\end{flalign}
The in-phase component of the output noise floor then contributes to the amplitude noise
and the quadrature component to the phase noise. These are well known linear mechanisms
that contribute to the active device amplitude and phase noise. But are there other
sources of noise that their contribution is
be higher than that of the linear mechanism? We know that flicker noise has higher levels
than the thermal noise or shot noise at low offset frequencies. The flicker noise can be
both in-phase- or quadrature-modulated by the carrier to higher frequencies. These
nonlinear processes then contribute to the active-device overall noise which exceeds the
output additive noise of the device. 

\begin{figure}[htb]
    \centering
    \ifusetikzpdf
    \includegraphics[]{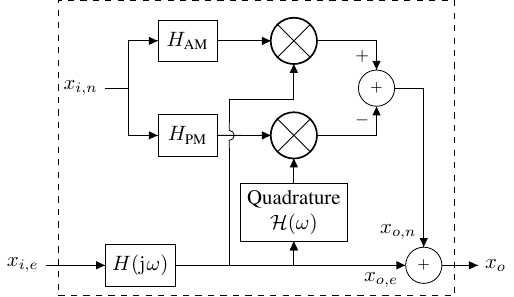}
    \else
    \tikzsetnextfilename{sch_nl_system}
    \input{./tikz/sch_nl_system.tex}
    \fi
    \caption{
        Model of active device incorporating AM and PM noise mechanisms.
}
    \label{fig:sch_nl_system}
\end{figure}

\subsection{Basic definitions} \label{sec:nl_basics}
In order to provide a foundation for analysis of nonlinear noise processes in active
devices, we use
the block diagram shown in Fig.~\ref{fig:sch_nl_system}. Later on, we will see that the
active devices we 
deal with follow the behavior modeled in this figure. The nonlinear device has two
inputs, the \emph{excitation} input $x_{i,e}$ and the \emph{noise} input $x_{i,n}$.
(the noise input is inside the device). The
block shown as $H(\ju\omega)$ models the linear part of the system. It gets an input and
linearly transforms it with a transfer function of $H(\ju\omega)$. Let's assume a single
tone input of 
\begin{flalign}\label{eq:basics_xie}
    x_{i,e} (t) =X_{1}\cos(\omega_0 t),
\end{flalign}
Where $A_{1}$ and $\omega_0$ are the amplitude and angular frequency of the excitation.
The output of this linear block is 
\begin{flalign}\label{eq:basics_xoe}
    x_{o,e} (t) = X_{o,1}\cos(\omega_0 t + \phi_1),
\end{flalign}
where
\begin{flalign}\label{eq:basics_xoe_mag_phase}
    \begin{dcases}
        X_{o,1} = X_1 |H(\ju\omega)| \\
        \phi_1 = \arg(H(\ju\omega))
    \end{dcases}.
\end{flalign}
Now lets look at the nonlinear part of the device modeling the AM and PM noise mechanisms.
The nonlinear AM part of the device gets a noise input, $x_{i,n}$. This noise
input passes through a linear block
with a transfer function of $H_\text{AM}$ and is modulated by the \emph{output} of the
linear part of the system, $x_{o,e}$. The nonlinear PM part of the device does something
similar. The noise first passes through a linear block, $H_\text{PM}$, and then is
modulated by the quadrature-transformed output of the $H(\ju\omega)$. The quadrature signal 
of $x_{o,e}$ is generated via the Hilbert transform block with a transfer
function of
\begin{flalign}\label{eq:basics_hilbert}
    \mathcal{H}(\omega) = -\ju\sign(\omega) = 
    \begin{cases}
        -\ju & \text{for} \quad \omega>0 \\
        +\ju & \text{for} \quad \omega<0
    \end{cases},
\end{flalign}
where $\mathcal{H}(\omega)$ denotes the Hilbert transform. When the input
of this block is an ideal monotone $X_{o,1}\cos(\omega_0 t+\phi_1)$, the output is 
$X_{o,1}\sin(\omega_0 t+\phi_1)$, without affecting the amplitude of the signal.
The output of the nonlinear noisy part of the device, $x_{o,n} (t)$, can therefore be 
written as
\begin{flalign}\label{eq:basics_xo}
    \nonumber
    x_{o,n} (t) ={}& X_{o,1} \big[h_\text{AM}(t)*x_{i,n}(t) \big]
    \cos(\omega_0 t + \phi_1) \\
    &- X_{o,1}\big[ h_\text{PM}(t)*x_{i,n}(t) \big]\sin(\omega_0 t + \phi_1)
\end{flalign}
where $*$ denotes convolution and $h_\text{AM}$ \& $h_\text{PM}$ are the inverse Fourier
transform of $H_\text{AM}$ \& $H_\text{PM}$, respectively.
The lower $i$, $o$, $e$ and $n$ indices denote the \emph{input}, \emph{output},
\emph{excitation} and \emph{noise}, respectively. We use this naming convention throughout
this paper repeatedly. The AM and PM noise of the output signal can consequently be found
using the relation given in \eqref{eq:basics_apn_anpn} as
\begin{flalign}\label{eq:basics_apn}
    a_n(t) = h_\text{AM}*x_{i,n}(t) \quad \text{and} \quad
    \phi_n(t) = h_\text{PM}*x_{i,n}(t),
\end{flalign}
Our convention to define $H_\text{AM}$ and $H_\text{PM}$ significantly simplifies the
nonlinear noise analysis. By deriving $H_\text{AM}$ and $H_\text{PM}$ and having the PSD
of the input noise process, The amplitude and phase noise of the output signal can be
found, or mathematically
\begin{flalign}\label{eq:basics_psd_apn}
    S_{a_n}(\omega) = |H_\text{AM}|^2 S_{x_{i,n}}(\omega)
    \ \ \text{and} \ \
    S_{\phi_n}(\omega) = |H_\text{PM}|^2 S_{x_{i,n}}(\omega).
\end{flalign}

In this paper, we mainly deal with nonlinear behaviors that both $H_\text{AM}$ and
$H_\text{PM}$ are real numbers and they have negligible variations with respect to
frequency. In this type of nonlinear elements we can write
\begin{flalign}\label{eq:basics_an}
    a_n(t) = H_\text{AM} x_{i,n}(t) \quad \text{and} \quad
    \phi_n(t) = H_\text{PM} x_{i,n}(t). 
\end{flalign}

Now that we have modeled a nonlinear device, we investigate some simple nonlinear
elements. We derive $H_\text{AM}$ and $H_\text{PM}$ for these elements. Our main objective
in this report to derive AM and PM transfer functions. In Section~\ref{sec:nl_rc}
we derive these transfer functions for a more complicated nonlinear device.

\subsection{Noise in memoryless nonlinear devices}
\label{sec:nl_dev_wom}
In this section, we look at the nonlinear behavior of elements whose output is a function
of instantaneous value of their inputs. We call this class of devices as memoryless, as
their output does not store any information about the input at any time earlier, hence the
naming.

\begin{figure}[htb]
    \centering
\subfloat[\label{fig:sch_nlg}]{
    \ifusetikzpdf
    \includegraphics[]{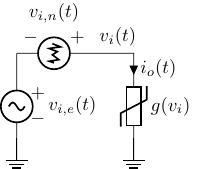}
    \else
    \tikzsetnextfilename{sch_nlg}
    \input{./tikz/sch_nlg.tex}
    \fi
} 
\subfloat[\label{fig:sch_nltc}]{
    \ifusetikzpdf
    \includegraphics[]{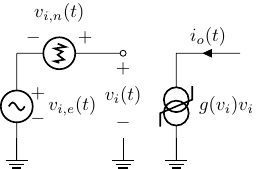}
    \else
    \tikzsetnextfilename{sch_nltc}
    \input{./tikz/sch_nltc.tex}
    \fi
}
    \caption{
        Sinusoidal stimulation of (a) nonlinear conductance and (b) nonlinear
        transconductance in presence of noise.
}
    \label{fig:sch_nlg_tc}
\end{figure}

\subsubsection{Nonlinear conductance and transconductance} 
Let's assume the nonlinear conductance or transconductance illustrated in
Fig.~\ref{fig:sch_nlg_tc}, whose output current is a function of the instantaneous value
of the input voltage
\begin{flalign}\label{eq:nl_dev_wom_fvi}
    i_o &= f(v_i) 
\end{flalign}
\begin{flalign}\label{eq:nl_dev_wom_taylor}
    i_o &= \sum_{k=0}^{\infty} \frac{1}{k!}f^{(k)}(0) v_i^k, 
\end{flalign}
where $f^{(k)}(v_i)$ denotes the $k$'th derivative of $f(v_i)$.
It is well known that coefficients $f_k$ are responsible for $k$'th order nonlinearity mechanism.
Now let's assume the dc current passing through the conductance and transconductance in the
absence of any excitation is zero, $f(0)=0$, and only consider the first and second order
coefficients. We define these terms as 
\begin{flalign}\label{eq:nl_dev_wom_g1g2}
    g_1 = f'(0), \quad \text{and} \quad
    g_2 = \frac12 f''(0).
\end{flalign}
The coefficient $g_1$ corresponds to a perfectly linear conductance and $g_2$ is
the first nonlinear coefficient and causes the current passing through the
element to be proportional to the square of the excitation voltage, causing a second
harmonic. Assuming only $g_1$ and $g_2$, the output current passing through the nonlinear
conductance can be written as
\begin{flalign}\label{eq:nl_dev_wom_g1g2_io}
    i_o \approx g_1v_i + g_2v_i^2.
\end{flalign}
Now we investigate the effect of second order nonlinearity on the output current of the
element in presence of input noise. The input voltage of the element, $v_i(t)$, has an
excitation part, $v_{i,e}(t)$, and an undesired noise part, $v_{i,n}(t)$, as
\begin{flalign}\label{eq:nl_dev_wom_vi}
    v_i(t) = v_{i,e}(t) + v_{i,n}(t).
\end{flalign}
We assume a single-tone excitation waveform of 
\begin{flalign}\label{eq:nl_dev_wom_vie}
    v_{i,e}(t) = V_1\cos(\omega_0 t)
\end{flalign}
where $V_1$ is the amplitude of the signal and $\omega_0$ is its angular frequency.
The device output current can then be written as
\begin{flalign}\label{eq:nl_dev_wom_io}
    \nonumber
    i_o(t) ={}& g_1 V_1\cos(\omega_0 t) + g_1v_{i,n}(t) 
    + g_2 V_1^2\cos^2(\omega_0 t) \\
    &+ 2g_2V_1 v_n(t) \cos(\omega_0 t) + g_2v_n^2(t) 
\end{flalign}
Now let's have a look at different terms of \eqref{eq:nl_dev_wom_io}. The first and second 
terms are linear transformation of the excitation and noise part of the input to output
current, respectively. The 3rd term contains the second harmonic of the signal and a dc terms
caused by the second order nonlinearity. The 4th term is the input noise modulated by the
input sinusoidal excitation and contributes to amplitude noise. Finally, the 5th term is
the noise self modulation term and therefore negligible.
Among these terms, we are only interested in the bandpass terms around $\omega_0$.
Throughout this paper, we only focus on nonlinear noise processes and neglect the linear
additive noise terms that are well modeled in the reference books \cite{Razavi2011}. With these
assumptions, the output terms around $\omega_0$ can be
written as
\begin{flalign}\label{eq:nl_dev_wom_io_e+n}
    i_o(t)\Big|_{\text{at }\omega_0} =  
    g_1V_1 \left[ 1 + 2\frac{g_2}{g_1} v_n(t) \right] \cos(\omega_0 t)
\end{flalign}
The noise term generated by the nonlinear device is in-phase with the carrier which
corresponds to amplitude noise. Therefore, we can write

\begin{flalign}\label{eq:nl_dev_wom_ham}
    H_\text{AM} = &  2\frac{g_2}{g_1},
\end{flalign}
The AM noise PSD can consequently be found using the basic relations in \eqref{eq:basics_psd_apn}.
Equation \eqref{eq:nl_dev_wom_io_e+n} does not contain any quadrature carrier and shows
that a memoryless nonlinear device is fundamentally unable to generate
phase noise. This is independent of degree of approximation in considering the Taylor
expansion coefficients in \eqref{eq:nl_dev_wom_taylor}. Even if we consider higher order terms 
$g_kv^{k}$ (where $g_k=f^{(k)}(0)/k!$) in the Taylor expansion series, these terms
cannot generate any
quadrature component. For instance, applying the input voltage in \eqref{eq:nl_dev_wom_vi}
gives an output current of 
\begin{flalign}\label{eq:nl_dev_wom_gk}
    g_kv^{k}_{i} = g_k\sum_{p=0}^{k} 
    \binom{k}{p} V^p_1\cos^p(\omega_0 t)v^{k-p}_{i,n}(t) .
\end{flalign}
None of the terms in the polynomial expansion of $g_kv^{k}_{i}$ can generate a
quadrature component since $\sin(\omega_0 t)$ is an odd function and $\cos(\omega_0 t)$ is
an even function
\begin{flalign}\label{eq:nl_dev_wom_nonquad}
    \int_{-\pi/\omega_0}^{\pi/\omega_0} \text{d}t [\cos(\omega_0 t)]^p \sin(\omega_0 t) = 0.
\end{flalign}
It is noteworthy that there are additional signal and noise bandpass terms around
$\omega_0$. For instance the 3rd and 5th order nonlinearity terms generate signal terms
that are proportional to
$V^3_1\cos(\omega_0 t)$ and $V^5_1\cos(\omega_0 t)$ that are responsible for gain
compression. Similarly, the 4th and 6th order nonlinearity terms generate AM noise terms
that are proportional to
$V^3_1v_{i,n}(t)\cos(\omega_0 t)$ and $V^5_1v_{i,n}(t)\cos(\omega_0 t)$ that depending on
the sign of $g_4$ and $g_6$ lead to AM noise compression or expansion. Throughout this
paper, we assume moderate excitation levels such that these higher order terms are well
below the linear and 2nd order nonlinear terms, or equivalently
\begin{flalign}\label{eq:nl_dev_wom_nonquad_fund}
    |g_{2m+1}V_1^{2m+1}| \ll |g_1V_1|,
\end{flalign}
and for the AM noise terms around $\omega_0$
\begin{flalign}\label{eq:nl_dev_wom_nonquad_2nd}
    |g_{2m+2}V_1^{2m+1}| \ll |g_2V_1|,
\end{flalign}
where $m$ is a positive non-zero integer.
It is noteworthy that the analysis provided here is also true for any memoryless nonlinear
system. This can be a system composed of \emph{only} nonlinear conductances. Both the
excitation and the desired output can also be a voltage or a current. If we have an
output-input relation of 
\begin{flalign}\label{eq:nl_dev_wom_general_xo}
    x_o = \alpha_1 x_i + \alpha_2 x_i^2 ,
\end{flalign}
the AM noise transfer function will be
\begin{flalign}\label{eq:nl_dev_wom_general_ham}
    H_\text{AM} = &  2\frac{\alpha_2}{\alpha_1}.
\end{flalign}

\begin{figure}[htb]
    \centering
\subfloat[\label{fig:sch_nlg_c}]{
    \ifusetikzpdf
    \includegraphics[]{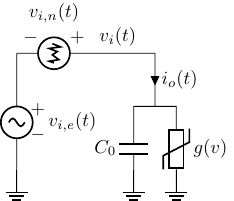}
    \else
    \tikzsetnextfilename{sch_nlg_c}
    \input{./tikz/sch_nlg_c.tex}
    \fi
} 
\subfloat[\label{fig:sch_nlc_g}]{
    \ifusetikzpdf
    \includegraphics[]{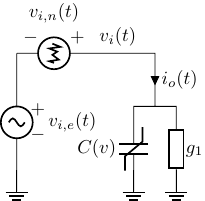}
    \else
    \tikzsetnextfilename{sch_nlc_g}
    \input{./tikz/sch_nlc_g.tex}
    \fi
} 
    \caption{
        Sinusoidal stimulation of (a) nonlinear conductance in parallel with capacitance and (b)
        nonlinear capacitance in parallel with conductance in presence of noise.
}
    \label{fig:sch_nl_gc}
\end{figure}

\subsection{Noise in nonlinear devices with memory}
\label{sec:nl_dev_wm}
In the previous section, we saw that a nonlinear memoryless element is fundamentally unable
to generate PM noise. For phase noise generation,
modulation of the baseband noise by the quadrature carrier, rather than in-phase carrier
in memoryless element, is necessary. In this section we introduce two simple element that
cause quadrature carrier modulation: linear capacitance in parallel with nonlinear
conductance, shown in Fig.~\ref{fig:sch_nlg_c}, and linear conductance in parallel with
nonlinear capacitance shown in Fig.~\ref{fig:sch_nlc_g}. Here we just
present the principle of quadrature-carrier noise modulation. Later on in
section~\ref{sec:nl_rc} we demonstrate more sophisticated circuits that we face in
practice. 

\subsubsection{Linear capacitance and nonlinear conductance}
Now let's have a look at the simple circuit shown in Fig.~\ref{fig:sch_nlg_c} composed of
a nonlinear
conductance and a linear capacitance. The circuit is excited by an external monotone
voltage source and a baseband noise is added to the excitation. We look at the behaviors
of the current
passing through the circuit and find the properties of modulated noise caused by the
nonlinearity of the conductance. The conductance has just a 2nd order nonlinearity 
with the governing equation as $g(v)=g_2v^2$.
The governing equation of the circuit can be written as
\begin{flalign}\label{eq:nl_dev_wm_nlg_gov}
    i_{o}(t) = C_0\dot{v}_{i}(t) + g_2v_i^2(t) ,
\end{flalign}
We also used the dot convention for the time derivative operation 
\begin{flalign}\label{eq:dot}
    \dot{v}(t) := \frac{\text{d}}{\text{d}t}v(t).
\end{flalign}
Assuming the excitation waveform and input noise to be same as 
\eqref{eq:nl_dev_wom_vi} and \eqref{eq:nl_dev_wom_vie},
the output current passing through the circuit can be written as
\begin{flalign}\label{eq:nl_dev_wm_nlg_io}
    \nonumber
    i_{o}(t) ={}& -C_0\omega_0V_1\sin(\omega_0) + C_0\dot{v}_{i,n}(t) 
    + g_2V_1^2\cos^2(\omega_0 t) \\
    &+ 2g_2V_1 v_{i,n}(t) \cos(\omega_0 t) + g_2v^2_{i,n}(t),
\end{flalign}
Now we look at the bandpass current and the modulated noise terms caused by the
nonlinearity. 
\begin{flalign}\label{eq:nl_dev_wm_nlg_io_des}
    \nonumber
    i_{o}(t) \Big|_{\text{at }\omega_0} 
    ={}& -C_0\omega_0V_1\sin(\omega_0 t) + 2g_2V_1 v_{i,n}(t) \cos(\omega_0 t) \\
    \nonumber
    ={}& C_0\omega_0V_1\cos(\omega_0 t+\pi/2) \\
    &+ 2g_2V_1 v_{i,n}(t) \sin(\omega_0 t+\pi/2) 
\end{flalign}
The modulated noise and the signal are in quadrature. Although the noise is, with respect
to input, in-phase modulated by the nonlinear conductance, the carrier signal itself is
quadrature-transformed by time derivation at the linear capacitance. The
PM noise transfer function can be written as 
\begin{flalign}\label{eq:nl_dev_wm_nlg_hpm}
    H_\text{PM} = -\frac{2g_2}{C_0\omega_0}
\end{flalign}

\subsubsection{Linear conductance and nonlinear capacitance}
Another circuit that causes phase noise generation is interaction between a linear
conductance and a first order nonlinear capacitance, illustrated in
Fig.~\subref*{fig:sch_nlc_g}. In order to analyze a circuit with nonlinear capacitor, we
need to derive its governing equation first. Let's assume a nonlinear capacitor with a
Taylor expansion series of 
\begin{flalign}\label{eq:nl_dev_cap_taylor}
    C(v) = \sum_{k=0}^{\infty} \frac{1}{k!}C^{(k)}(0) v_i^k, 
\end{flalign}
The current passing through the capacitor terminals can be written as
\begin{flalign}\label{eq:nl_dev_cap_ic_product}
    \nonumber
    i_C &= \frac{\text{d}Q}{\text{d}t} 
    = \frac{\text{d}\big(C(v)\times v \big)}{\text{d}t} \\
    &= \big[ C(v) + C'(v)\times v \big] \frac{\text{d}v}{\text{d}t}
\end{flalign}
Where we used the chain and product derivative rules. Substituting $C(v)$ and its derivative
into \eqref{eq:nl_dev_cap_taylor}, the governing equation of the capacitor can be
written as 
\begin{flalign}\label{eq:nl_dev_cap_ic_taylor}
    i_C = \left[ \sum_{k=0}^{\infty} \frac{k+1}{k!}C^{(k)}(0) v_i^k \right] 
    \frac{\text{d}v}{\text{d}t}.
\end{flalign}
Now we define the linear and 2nd-order nonlinear capacitance coefficients as 
\begin{flalign}\label{eq:nl_dev_cap_c0c1}
    C_0 = C(0) \quad \text{and} \quad C_1 = 2C'(0).
\end{flalign}
The governing equation of the capacitor considering only the linear coefficient and
2nd-order nonlinearity can be written as
\begin{flalign}\label{eq:nl_dev_cap_gov}
    i_C(t) = (C_0 + C_1v)\dot{v}(t) .
\end{flalign}
Now let's consider the circuit shown in Fig.~\ref{fig:sch_nlc_g} and assume the
capacitance to be purely nonlinear of second order, $C_0 = 0$.
The output current of the circuit can then be formulated as
\begin{flalign}\label{eq:nl_dev_wm_nlc_gov}
    i_{o}(t) = g_1v + C_1v_{i}(t)\dot{v}_{i}(t).
\end{flalign}
Assuming the excitation waveform and input noise to be same as 
\eqref{eq:nl_dev_wom_vi} and \eqref{eq:nl_dev_wom_vie},
the output current passing through the circuit can be written as
\begin{flalign}\label{eq:nl_dev_wm_nlc_io}
    \nonumber
    i_{o}(t) ={}& g_1V_1\cos(\omega_0) + g_1v_{n,i}(t)
    - C_1\omega_0V_1^2\sin(\omega_0 t)\cos(\omega_0 t) \\
    \nonumber
    &- C_1\omega_0V_1v_{i,n}(t)\sin(\omega_0) 
    + C_1\omega_0V_1\dot{v}_{i,n}(t)\cos(\omega_0) \\
    &+ v_{i,n}(t)\dot{v}_{i,n}(t) .
\end{flalign}
Once again, we only look at the signal and modulated noise terms around $\omega_0$.
\begin{flalign}\label{eq:nl_dev_wm_nlc_io_des}
    \nonumber
    i_{o}(t) \Big|_{\text{at }\omega_0} ={}&
    g_1V_1\cos(\omega_0) - C_1\omega_0V_1v_{i,n}(t)\sin(\omega_0) \\
    &+ C_1\omega_0V_1\dot{v}_{i,n}(t)\cos(\omega_0) .
\end{flalign}
The first term on the right-hand side of this equation is the desired current 
that passes through the linear conductance. The second term is the baseband noise
modulated by the quadrature carrier and leads to phase noise generation. The last term
is the derivative of the baseband noise modulated by the in-phase carrier. It is
noteworthy that the derivative of the baseband noise leads to small AM noise power
spectral densities at close-in carrier offset frequencies. It is a legitimate concern that
this derivative operation leads to a high noise PSD at high frequencies, but in practical
circuits a series resistance prevents a purely derivative operation on the input noise.
Therefore, the output current mainly contains PM noise with a transfer function of 
\begin{flalign}\label{eq:nl_dev_wm_nlc_hpm}
    H_\text{PM} = \frac{C_1\omega_0}{g_1}
\end{flalign}

\begin{figure}[htb]
    \centering
    \ifusetikzpdf
    \includegraphics[]{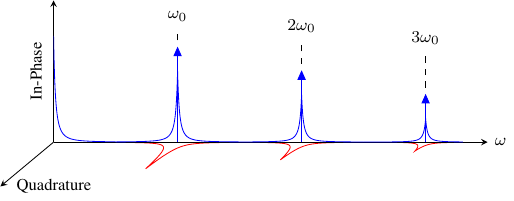}
    \else
    \tikzsetnextfilename{sch_rc_out_spec}
    \input{./tikz/sch_rc_out_spec.tex}
    \fi
    \caption{
        Output spectrum of a nonlinear circuit showing both in-phase and quadrature noise.
    }
    \label{fig:sch_rc_out_spec}
\end{figure}

\section{Noise in nonlinear RC circuits}\label{sec:nl_rc}
So far we have seen how nonlinear conductance leads to AM noise generation and a
combination of nonlinearity of conductance and capacitance causes PM noise generation.
While these principles are simple and easy to understand, unfortunately, real examples of
AM and PM noise require solving more complicated nonlinear differential equations.
The nonlinear nature of these circuits causes many harmonics to be generated, illustrated
in Fig.~\ref{fig:sch_rc_out_spec} as Dirac delta pulse arrows in frequency domain. The
harmonics at various nodes of the circuit are also mixed with the baseband noise
processes in the circuit. Depending on the nature of the mixing process, as described
in the previous section, the generated modulated noise can be in-phase or in
quadrature (or both) with a given harmonic. Therefore, the noise has
two components, shown in Fig.~\ref{fig:sch_rc_out_spec} with two separate axes. These
noise side-bands at different frequencies are in balance via the nonlinear governing
equation of the circuit. The linear mechanisms, existing in linear resistor, capacitor
and inductor, cannot change the frequency of these noise side-bands, but can leak the
in-phase component of a noise sideband to the quadrature component at the same frequency
or vice versa. The nonlinear components of the circuit, however, change the frequency of
a noise sideband. This change in frequency can also be within quadrature or in-phase
transformation of the noise sideband. The balance between these noise side-bands are also
determined by the governing nonlinear differential equation (or set of equations) of the
circuit. This type of differential equations usually does not have a closed-form solution.
It could be argued that the only way to find a response is through the use of computer
aided design (CAD) tools. The downside of this approach to find the optimum point is that
we close our eyes to the
dynamics of the system. An alternative approach is to find an approximate solution to these
equations that model the behavior of the system parametrically. This approach helps the 
designer to both understand which parameter to vary and move in the right direction for
the optimization. 

In this section, we analyze three circuits with nonlinear conductance and capacitance.
We show that with some reasonable assumptions their response can be found with very good
precision. Later on, in section~\ref{sec:bipolar} we directly apply the results of this
discussion to bipolar transistor and find an expression for the output AM and PM noise.

\subsection{Noise in RC circuit with nonlinear conductance}
\label{sec:rc_nlg}
\begin{figure}[htb]
    \centering
    \ifusetikzpdf
    \includegraphics[]{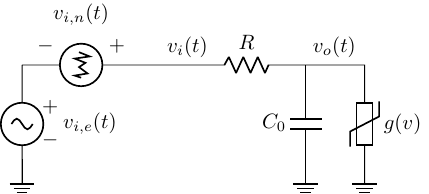}
    \else
    \tikzsetnextfilename{sch_rc_nlg}
    \input{./tikz/sch_rc_nlg.tex}
    \fi
    \caption{
       Nonlinear RC circuit with a nonlinear conductance.
}
    \label{fig:sch_rc_nlg}
\end{figure}

Let's consider the nonlinear series RC circuit shown in Fig.~\ref{fig:sch_rc_nlg} composed
of a linear resistor and a capacitor in parallel with a purely 2nd-order nonlinear
conductance.  The conductance has just a 2nd order nonlinearity 
as $g(v)=g_2v^2$.

%
We assume the excitation waveform and input noise to be same as 
\eqref{eq:nl_dev_wom_vi} and \eqref{eq:nl_dev_wom_vie}.
The governing equation of the circuit can be written as
\begin{flalign}\label{eq:sec_rc_nlg_gov}
    v_{o}(t) = v_{i}(t) - RC_0 \dot{v}_{o}(t) - g_2Rv_o^2(t) ,
\end{flalign}
Equation \eqref{eq:sec_rc_nlg_gov} has a $g_2Rv_{o}^2(t)$ term which makes it a nonlinear
first order differential equation.
In order to find an expression for the output voltage $v_o(t)$, we decompose the response
of the circuit to the \emph{excitation} term $v_{o,e}(t)$ and the \emph{noise}
$v_{o,n}(t)$ term as
\begin{flalign}\label{eq:sec_rc_nlg_vo_e+n}
    v_{o}(t) = v_{o,e}(t) + v_{o,n}(t).
\end{flalign}
Note that the governing equation of the circuit in \eqref{eq:sec_rc_nlg_gov} is nonlinear
and we cannot simply use the superposition principle that is widely
used in linear circuits. We expect the noise response $v_{o,n}(t)$ to be dependent upon the
excitation voltage. First we assume $v_{i,n}(t)=0$ and estimate the excitation part of the
response. The differential equation in this case can simply be written as
\begin{flalign}\label{eq:sec_rc_nlg_voe_gov}
    v_{o,e}(t) = v_{i,e}(t) - RC_0 \dot{v}_{o,e}(t) 
    - g_2Rv_{o,e}^2(t) .
\end{flalign}
The nonlinear nature of this differential equation leads to generation of harmonics of the
excitation waveform. Therefore, the steady-state response has a form of
\begin{flalign}\label{eq:sec_rc_nlg_vok}
    v_{o,e}(t) = 
    \sum_{k=0}^{\infty} v_{o,k}(t) =
    \sum_{k=0}^{\infty} V_{o,k} \cos(k\omega_0 t + \phi_k)
\end{flalign}
where $V_{o,k}$ and $\phi_k$ are the amplitude and the offset phase of the $k$'th
harmonic. Now we try to find an approximate solution for the response in presence of single
tone stimulation of the circuit. Considering only $k=0,1$, and substituting
\eqref{eq:nl_dev_wom_vie} and \eqref{eq:sec_rc_nlg_vok} in \eqref{eq:sec_rc_nlg_voe_gov}
and neglecting the second and higher order harmonics, a first order approximation of the
response can found as
\begin{flalign}\label{eq:sec_rc_nlg_vo1}
    v_{o,e}(t) \Big|_{\substack{\text{1st} \\ \text{harm.} }} 
    \approx  V_{o,1}\cos(\omega_0 t +\phi_1) ,
\end{flalign}
where 
\begin{flalign}\label{eq:sec_rc_nlg_vo1_mag_phi}
    \begin{dcases}
        H(\ju\omega) = \frac{1}{1+\ju RC_0\omega_0} \\
        V_{o,1} = |H(\ju\omega_0)|V_{1} \\
        \phi_1 = \arg \big( H(\ju\omega_0) \big)
    \end{dcases}.
\end{flalign}
This is indeed the response of the circuit in the absence of any nonlinearity, $g(v)=0$.
A second order approximation considering $k=0,1,2$ can be performed to find the second
harmonic. Since $v_{i,e}(t)$ does not contain any harmonics, the term
$g_2RV^2_{o,1}\cos{\omega_0 t}$ acts as a \emph{source term} for the second harmonic response. 
This new source term has a dc component and a second harmonic as
\begin{flalign}\label{eq:sec_rc_nlg_vo2_rec}
    -g_2R \left[ 
    v_{o,e}(t) \Big|_{\substack{\text{1st} \\ \text{harm.} }} 
    \right]^2
    = -\frac12 g_2RV^2_{o,1}[1+\cos(2\omega_0 t +2\phi_1)]
\end{flalign}
Therefore the new output voltage approximation contains a dc term and a second harmonic as
\begin{flalign}\label{eq:sec_rc_nlg_vo2_mag_phi}
    \begin{dcases}
        V_{o,0} = -\frac12 g_2RV^2_{o,1} \\
        V_{o,2} = \frac12 g_2R \big| H(\ju2\omega_0)\big | V^2_{o,1} \\
        \phi_2 = 2\phi_1 + \arg \big( H(\ju2\omega_0) \big) + \pi
    \end{dcases}.
\end{flalign}
In every step to find a better approximate response, new mixing terms appear. These mixing
terms caused by nonlinearity are proportional to $g_2RV_{o,1}$. The magnitude of the
$k$'th harmonic therefore is proportional to $(g_2RV_{o,1})^k$ and higher order terms.
The convergence of the
response depends upon both the excitation amplitude and its frequency and requires further
analysis. For the rest of our analysis we assume
moderate excitation levels $|g_2RV_{o,1}| \ll 1$ and approximate the output with its first
harmonic given in \eqref{eq:sec_rc_nlg_vo1}. 

%

\begin{figure}[htb]
    \centering
    \ifusetikzpdf
    \includegraphics[]{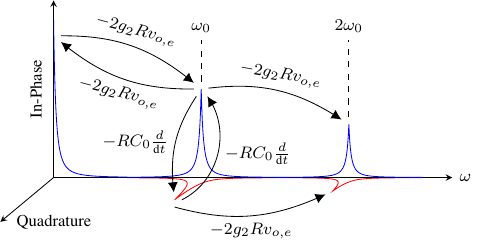}
    \else
    \tikzsetnextfilename{sch_rc_nlg_mechanism}
    \input{./tikz/sch_rc_nlg_mechanism.tex}
    \fi
    \caption{
        Formation of noise side-bands in $RC$ circuit with nonlinear conductance.}
    \label{fig:sch_rc_nlg_mechanism}
\end{figure}

Now we derive the governing differential equation of the output noise
voltage by substituting \eqref{eq:nl_dev_wom_vi} and \eqref{eq:sec_rc_nlg_vo_e+n} into
governing differential equation of the circuit in \eqref{eq:sec_rc_nlg_gov}.
This gives output noise voltage differential equation as 
\begin{flalign}\label{eq:sec_rc_nlg_von_gov}
    \nonumber
    v_{o,n}(t) ={}& v_{i,n}(t) - RC_0 \dot{v}_{o,n}(t) - 2g_2Rv_{o,e}(t)v_{o,n}(t) \\
    & - g_2Rv_{o,n}^2(t).
\end{flalign}
The first 2 terms on the right-hand side of \eqref{eq:sec_rc_nlg_von_gov} are responsible
for noise shaping in linear
systems. The 3rd term is caused by the nonlinearity of conductance and in-phase-modulates
the noise around the carrier. The 4th term is noise self
modulation and negligible, as the input and --- consequently ---- output noise voltages
are small. There is another
effect that is obscure in the 2nd term of \eqref{eq:sec_rc_nlg_von_gov}. Having the output
noise in-phase modulated
by the nonlinear conductance, the derivative of this AM noise causes PM noise
generation. Linear and nonlinear transformation between different noise components are
illustrated graphically in Fig.~\ref{fig:sch_rc_nlg_mechanism}

Now we derive different components of the output noise voltage. By neglecting the
noise-self-modulation term in \eqref{eq:sec_rc_nlg_von_gov}, the output noise
differential equation can be rewritten as 
\begin{flalign}\label{eq:sec_rc_nlg_von_gov_aprx}
    v_{o,n}(t) \approx{}& v_{i,n}(t) - RC_0 \dot{v}_{o,n}(t) - 2g_2Rv_{o,e}(t)v_{o,n}(t).
\end{flalign}
This equation is linear ordinary differential equation and has an exact solution. However,
since $v_{o,n}(t)$ has a time-dependent coefficient, $v_{o,e}(t)$, the general form of the
response is complicated. Therefore, we try to estimate its response directly from the
\eqref{eq:sec_rc_nlg_von_gov_aprx}.
The output noise $v_{o,n}(t)$ has both in-phase and quadrature sidebands at the harmonics
of the excitation frequency. These terms
are tightly coupled via various terms on the right-hand side of
\eqref{eq:sec_rc_nlg_von_gov_aprx}. The second term 
$RC_0 \dot{v}_{o,n}(t)$ contains time derivative of the noise. This time-derivative does
not affect the baseband noise shaping, as the bandwidth of the input noise is in the
megahertz range and the linear cut-off frequency of the circuit is in the gigahertz range
that is approximately 3 orders of magnitude higher than the input noise frequencies. The
time derivative operation, however, couples the in-phase component of noise into
the quadrature component and vice versa. The 4th term modulates the baseband noise by the
carrier and is the source of AM noise generation in the first place.

\begin{table*}[t]
\caption{Detailed terms in decomposition of the output noise of RC circuit with nonlinear
    conductance into baseband, in-phase and quadrature components}
\centering
\renewcommand{\arraystretch}{1.3}
    \begin{tabular}{C{2.7cm}C{2.4cm}C{3.9cm}C{3.9cm}} 
    \hline
        term in \eqref{eq:sec_rc_nlg_von_gov_aprx} &
        baseband part & 
        in-phase part & 
        quadrature part  \\ 
    \hline\hline 
        $ v_{o,n}(t) $ & 
        $ v_{n,\textbf{BB}}(t) $ & 
        $ v_{n,\textbf{I}}(t)\cos(\omega_0t+\phi_1) $ & 
        $ v_{n,\textbf{Q}}(t)\sin(\omega_0t+\phi_1) $  \\
    \hline 
        $v_{i,n}(t)$ &  
        $v_{i,n}(t)$ &  
        --  & 
        --  \\
    \hline 
        $ - RC_0 \dot{v}_{o,n}(t) $ & 
        $ - RC_0 \dot{v}_{n,\textbf{BB}}(t)$ & 
        \begin{equation*}\begin{gathered}
                - RC_0 \dot{v}_{n,\textbf{I}}(t)\cos(\omega_0t+\phi_1) \\
                - RC_0\omega_0 v_{n,\textbf{Q}}(t)\cos(\omega_0t+\phi_1)
        \end{gathered}\end{equation*} &
        \begin{equation*}\begin{gathered}
                - RC_0 \dot{v}_{n,\textbf{Q}}(t)\sin(\omega_0t+\phi_1) \\
                + RC_0\omega_0 v_{n,\textbf{I}}(t)\sin(\omega_0t+\phi_1)
        \end{gathered}\end{equation*} \\
    \hline 
        $ -2g_2Rv_{o,e}(t)v_{o,n}(t)$ & 
        $ -g_2RV_{o,1}v_{n,\textbf{I}}(t) $ & 
        $ -2g_2RV_{o,1}v_{n,\textbf{BB}}\cos(\omega_0 t+\phi_1) $ & 
        --  \\
    \hline
\end{tabular}
\label{eq:sec_rc_nlg_terms}
\end{table*}

Now we estimate different noise components. We neglect the modulated noise around
the harmonics and approximate $v_{o,n}(t)$ to have three terms: 
the baseband noise $v_{n,\textbf{BB}}(t)$, 
the modulated in-phase component $v_{n,\textbf{I}}(t)$, 
and the modulated quadrature component
$v_{n,\textbf{Q}}(t)$. The output noise can then be formulated as
\begin{flalign}\label{eq:sec_rc_nlg_von_bbiq}
    \nonumber
    v_{o,n}(t) \approx{}& v_{n,\textbf{BB}}(t) 
    + v_{n,\textbf{I}}(t)\cos(\omega_0t+\phi_1) \\
    &+ v_{n,\textbf{Q}}(t)\sin(\omega_0t+\phi_1)
\end{flalign}
Substituting this equation in \eqref{eq:sec_rc_nlg_von_gov_aprx} and neglecting the terms
at the second harmonic, the
relation between these terms and input noise can be found. The result of this substitusion
for each term of \eqref{eq:sec_rc_nlg_von_gov_aprx} is shown in
Table~\ref{eq:sec_rc_nlg_terms}. Neglecting the noise terms modulated around the 2nd
harmonic, caused by the
nonlinearity of the system, we get a set of three \emph{linear and homogeneous} differential
equations describing the relation between different noise components: the basebad noise,
the in-phase-modulated noise and the quadrature-modulated noise. For the quadrature PM
noise we have
\begin{flalign}\label{eq:sec_rc_nlg_vnq_gov}
    v_{n,\textbf{Q}}(t) + RC_0\dot{v}_{n,\textbf{Q}}(t) 
    = RC_0\omega_0v_{n,\textbf{I}}(t).
\end{flalign}
This equation shows that the source of quadrature PM noise in the circuit is the in-phase
AM noise. This process has also been explained before, that this happens by the linear
time-derivative operation via the capacitor. For the in-phase AM noise we have
\begin{flalign}\label{eq:sec_rc_nlg_vni_gov}
    v_{n,\textbf{I}}(t) &+ RC_0\dot{v}_{n,\textbf{I}}(t) =
    - 2g_2RV_{o,1}v_{n,\textbf{BB}}(t)
    - RC_0\omega_0v_{n,\textbf{Q}}(t) .
\end{flalign}
The first term in the right-hand side of \eqref{eq:sec_rc_nlg_vni_gov} once again shows
the source of AM noise is the nonlinear conductance present in the
circuit. The second term in the right-hand side of \eqref{eq:sec_rc_nlg_vni_gov} shows that
the PM noise caused by the AM noise in the first
place, couples into AM noise by the time derivative operation at the capacitor. 

Finally, for the baseband noise we have
\begin{flalign}\label{eq:sec_rc_nlg_vnbb_gov}
    v_{n,\textbf{BB}}(t) + RC_0\dot{v}_{n,\textbf{BB}}(t) 
     = v_{i,n}(t) - g_2RV_{o,1}v_{n,\textbf{I}}(t).
\end{flalign}
The first term on the right hand side of this equation shows the main source of output
baseband noise is the input baseband noise. 
The second term is the AM noise folded back to baseband due to nonlinear conductance.
Although this set of equations, 
\eqref{eq:sec_rc_nlg_vnq_gov} to 
\eqref{eq:sec_rc_nlg_vnbb_gov}, are linear with a closed form analytical solution, we make
two additional approximations to simplify the result:

\begin{enumerate}
    \item We assume $g_2RV_{o,1} \ll 1$ and neglect the folding of the AM
        noise to baseband. 
    \item We assume the high frequency dynamics of the RC circuit, namely its linear time
        constant, does not shape the baseband, in-phase-modulated and quadrature-modulated
        noise. This assumption is equivalent to neglecting all time-derivatives in
        \eqref{eq:sec_rc_nlg_vnq_gov} to \eqref{eq:sec_rc_nlg_vnbb_gov}.
        Among all our approximations so far, this is the least important one and
        has almost \emph{no} effect on the solution. For instance, a lowpass filter with a
        bandwidth of \si{1~GHz} has a close-to-zero effect on the magnitude and phase of a
        low frequency noise with a maximum frequency of \si{1~MHz}.
\end{enumerate}

\begin{figure}[htb]
    \centering
\subfloat[\label{fig:plot_nlg_ham}]{
    \ifusetikzpdf
    \includegraphics[]{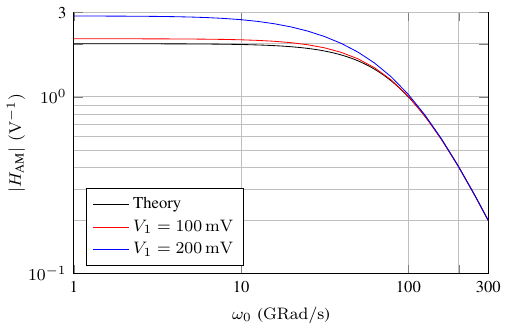}
    \else
    \tikzsetnextfilename{plot_nlg_ham}
    \input{./tikz/plot_nlg_ham.tex}
    \fi
}\\
\subfloat[\label{fig:plot_nlg_hpm}]{
    \ifusetikzpdf
    \includegraphics[]{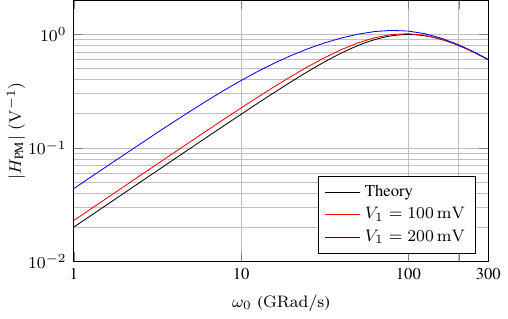}
    \else
    \tikzsetnextfilename{plot_nlg_hpm}
    \input{./tikz/plot_nlg_hpm.tex}
    \fi
}
    \caption
    {
    AM and PM noise transfer functions in RC circuit with nonlinear transconductance;
     $R=100~\si{\Omega}$, $C_0=100~\si{fF}$, $g_2=10~\si{mS/V}$, 
 }
\label{fig:plot_nlg}
\end{figure}

With these assumptions, we can write
\begin{subequations}
\begin{align}
    v_{n,\textbf{BB}}(t) &\approx v_{i,n}(t),
\label{eq:sec_rc_nlg_vnbb}\\
    v_{n,\textbf{I}}(t) &\approx 
    \frac{-2g_2R} {1+(RC_0\omega_0)^2} V_{o,1}v_{i,n}(t),
\label{eq:sec_rc_nlg_vni}\\
    v_{n,\textbf{Q}}(t) &\approx 
    \frac{-2g_2R^2C_0\omega_0} {1+(RC_0\omega_0)^2} V_{o,1}v_{i,n}(t)
\label{eq:sec_rc_nlg_vnq}
\end{align}
\end{subequations}
Although our analysis has been complicated, we ended up with elegant and simple results.
As the final step, we write the AM and PM noise transfer functions as
\begin{flalign}\label{eq:sec_rc_nlg_hamhpm}
    H_\text{AM} = \frac{-2g_2R} {1+(RC_0\omega_0)^2} \quad \text{and} \quad
    H_\text{PM} = \frac{2g_2R^2C_0\omega_0} {1+(RC_0\omega_0)^2}.
\end{flalign}
Fig.~\ref{fig:plot_nlg} shows the simulation results of an RC circuit with nonlinear
conductance in comparison with theory. The simulation was performed using Virtuoso analog design
environment commercial software from Cadence Design Systems, Inc. At the excitation level
of $V_1 = 100~mV$, the simulation results have a good matching with theory. As the
excitation level increases, higher order harmonics grow and their contribution to AM/PM
noise leads to deviation of simulation results from our analysis.

\begin{figure}[htb]
    \centering
    \ifusetikzpdf
    \includegraphics[]{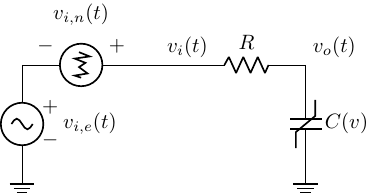}
    \else
    \tikzsetnextfilename{sch_rc_nlc}
    \input{./tikz/sch_rc_nlc.tex}
    \fi
    \caption{
       Nonlinear RC circuit with a nonlinear capacitance.
}
    \label{fig:sch_rc_nlc}
\end{figure}

\subsection{Noise in RC circuit with nonlinear capacitance}
\label{sec:rc_nlc}
A well known nonlinear storage element is a nonlinear capacitor. This nonlinearity
arises in active devices by the charge dependent length of depletion region in
bipolar transistors or dependence of gate-channel capacitance to the gate voltage in MOS
devices. In this section, we analyze a simple RC network with a nonlinear capacitance,
illustrated in Fig.~\ref{fig:sch_rc_nlc}. We assume a second-order nonlinear capacitance as
\begin{flalign}\label{eq:sec_rc_nlc_cv}
    C(v) = C_0 + C_1v.
\end{flalign}
The governing equation of the circuit in Fig.~\ref{fig:sch_rc_nlc} can be written as
\begin{flalign}\label{eq:sec_rc_nlc_gov}
    v_{o}(t) = v_{i}(t) - RC_0 \dot{v}_{o}(t) 
    - RC_1v_{o}(t) \dot{v}_{o}(t),
\end{flalign}
This equation has a $RC_1v_{o}(t) \dot{v}_{o}(t)$ term which makes it a nonlinear first
order differential equation. Once again, this equation does not have an analytical closed
form solution and we need to make some approximations to find a reasonable form for the
response.

In order to find an expression for the output voltage $v_o(t)$, once again, we divide the
response to two parts, the excitation part $v_{o,e}(t)$ and the noise part $v_{o,n}(t)$.
First we assume $v_{i,n}(t)=0$ and find the excitation part of the response. This gives us
the differential equation of
\begin{flalign}\label{eq:sec_rc_nlc_voe_gov}
    v_{o,e}(t) = v_{i,e}(t) - RC_0 \dot{v}_{o,e}(t) 
    - RC_1v_{o,e}(t) \dot{v}_{o,e}(t).
\end{flalign}
Once again, the nonlinear nature of this differential equation leads to generation of
harmonics of the
excitation waveform. Therefore, The steady-state response has the form given in previous
section in \eqref{eq:sec_rc_nlg_vok}. Considering only $k=0,1$, the response has the form
of linear RC circuit
given \eqref{eq:sec_rc_nlg_vo1} and \eqref{eq:sec_rc_nlg_vo1_mag_phi}.
A second order approximation and taking into account $k=2$ can be performed by
substituting the linear response in \eqref{eq:sec_rc_nlg_vo1} in the right-hand side of 
\eqref{eq:sec_rc_nlc_voe_gov}.
This substitution generates in new \emph{source term} for the second harmonic as
\begin{flalign}\label{eq:sec_rc_nlc_vo2_rec}
    - RC_1 \Big[ 
    v_{o,e}(t) \dot{v}_{o,e}(t) 
    \Big]_{\substack{\text{1st} \\ \text{harm.} }}
    = \frac12 RC_1\omega_0V_{o,1}^2\sin(2\omega_0t + 2\phi_1).
\end{flalign}
Therefore, the magnitude and phase of the 2nd harmonic can be approximated as
\begin{flalign}\label{eq:sec_rc_nlc_vo2_mag_phi}
    \begin{dcases}
        V_{o,2} = \frac12 RC_1\omega_0 \big| H(\ju2\omega_0)\big | V^2_{o,1} \\
        \phi_2 = 2\phi_1 + \arg \big( H(\ju2\omega_0) \big) - \frac12 \pi
    \end{dcases}.
\end{flalign}

This process can be done recursively to find the exact form of the steady state solution.
Substituting the $k$th order approximation of the response to the right-hand side of
\eqref{eq:sec_rc_nlc_voe_gov}
generates new sinusoidal terms that are one order more proportional to
$RC_1V_{o,1}\omega_0$. 
The harmonics of the output signal are therefore directly proportional to excitation
frequency. As the excitation frequency increases, the harmonics, on the one hand, grow
stronger because of the term $RC_1v_{o,e}(t) \dot{v}_{o,e}(t)$ in the governing equation,
and on the other hand, are shaped by the linear transfer function of the system. 
Here we avoid further discussion of convergence criteria and, similar to previous
discussion in analysis of RC circuit with nonlinear conductance, 
assume $RC_1V_{o,1}\omega_0 \ll 1$ and approximate the response with the
first order approximation given 
\eqref{eq:sec_rc_nlg_vo1} and \eqref{eq:sec_rc_nlg_vo1_mag_phi}.

\begin{figure}[htb]
    \centering
    \ifusetikzpdf
    \includegraphics[]{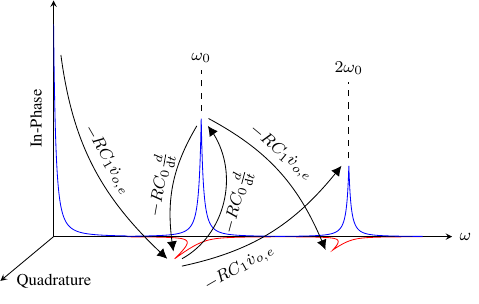}
    \else
    \tikzsetnextfilename{sch_rc_nlc_mechanism}
    \input{./tikz/sch_rc_nlc_mechanism.tex}
    \fi
    \caption{
        Formation of noise side-bands in $RC$ circuit with nonlinear capacitance.}
    \label{fig:sch_rc_nlc_mechanism}
\end{figure}

Now we derive the governing differential equation of the output noise
voltage by substituting \eqref{eq:nl_dev_wom_vie} and \eqref{eq:sec_rc_nlg_vo_e+n} into
governing differential equation of the circuit in \eqref{eq:sec_rc_nlc_gov}.
\begin{flalign}\label{eq:sec_rc_nlc_von_gov}
    \nonumber
    v_{o,n}(t) ={}& v_{i,n}(t) - RC_0 \dot{v}_{o,n}(t) 
    - RC_1v_{o,e}(t) \dot{v}_{o,n}(t) \\
    &- RC_1\dot{v}_{o,e}(t) v_{o,n}(t) 
    - RC_1v_{o,n}(t) \dot{v}_{o,n}(t).
\end{flalign}
The last term in the right-hand side of \eqref{eq:sec_rc_nlc_von_gov} is the noise self
modulation term and can be neglected. This approximation gives the noise linear ordinary
differential equation of
\begin{flalign}\label{eq:sec_rc_nlc_von_gov_aprx}
    \nonumber
    v_{o,n}(t) ={}& v_{i,n}(t) - RC_0 \dot{v}_{o,n}(t) 
    - RC_1v_{o,e}(t) \dot{v}_{o,n}(t) \\
    &- RC_1\dot{v}_{o,e}(t) v_{o,n}(t).
\end{flalign}
Now we look at the behavior of different terms of this equation. The first two terms are
responsible for
the noise shaping in linear systems. Once again, the second term has also another effect:
it transforms the quadrature noise into in-phase noise and vice versa. The third term
modulates the
derivative of the output noise by the carrier in-phase component. 
The derivatives of baseband noise are negligible at low offset frequencies (relative to
carrier frequency). Therefore,
with respect to the modulation of baseband noise, the third term is not expected to be
the main contributor to noise modulation. The 4th term, however, quadrature-modulates the
baseband noise and acts as a source term for the quadrature noise. These mechanisms are
also illustrated graphically in Fig.~\ref{fig:sch_rc_nlc_mechanism}.

\begin{table*}[t]
\caption{Detailed terms in decomposition of the output noise of RC circuit with nonlinear
    capacitance into baseband, in-phase and quadrature components}
\centering
\renewcommand{\arraystretch}{1.3}
\begin{tabular}{C{3.0cm}C{4.0cm}C{4.2cm}C{4.5cm}}
    \hline 
        term in \eqref{eq:sec_rc_nlc_von_gov_aprx} &
        baseband part & 
        in-phase part & 
        quadrature part \\
    \hline\hline 
        $ v_{o,n}(t) $ & 
        $ v_{n,\textbf{BB}}(t) $ & 
        $ v_{n,\textbf{I}}(t)\cos(\omega_0t+\phi_1) $ & 
        $ v_{n,\textbf{Q}}(t)\sin(\omega_0t+\phi_1) $ \\
    \hline 
        $v_{i,n}(t)$ &  
        $v_{i,n}(t)$ &  
        --  & 
        --  \\
    \hline 
        $ - RC_0 \dot{v}_{o,n}(t) $ & 
        $ - RC_0 \dot{v}_{n,\textbf{BB}}(t)$ & 
        \begin{equation*}\begin{gathered}
                - RC_0 \dot{v}_{n,\textbf{I}}(t)\cos(\omega_0t+\phi_1) \\
                - RC_0\omega_0 v_{n,\textbf{Q}}(t)\cos(\omega_0t+\phi_1)
            \end{gathered}\end{equation*} &
        \begin{equation*}\begin{gathered}
                -  RC_0 \dot{v}_{n,\textbf{Q}}(t)\sin(\omega_0t+\phi_1) \\
                + RC_0\omega_0 v_{n,\textbf{I}}(t)\sin(\omega_0t+\phi_1)
        \end{gathered}\end{equation*} \\
    \hline 
        \begin{equation*}\begin{gathered}
            - RC_1\frac{\text{d}}{\text{d}t}\left(v_{o,e}(t)v_{o,n}(t)\right)
        \end{gathered}\end{equation*} &
        \begin{equation*}\begin{gathered}
            -\frac12 RC_1V_{o,1}\dot{v}_{n,\textbf{I}}(t)
        \end{gathered}\end{equation*} &
        \begin{equation*}\begin{gathered}
            - RC_1V_{o,1}\dot{v}_{n,\textbf{BB}}(t)\cos(\omega_0 t+\phi_1)
        \end{gathered}\end{equation*} &
        \begin{equation*}\begin{gathered}
            RC_1\omega_0 V_{o,1}v_{n,\textbf{BB}}(t)\sin(\omega_0 t+\phi_1)
        \end{gathered}\end{equation*} \\
    \hline
\end{tabular}
\label{eq:sec_rc_nlc_terms}
\end{table*}

Once again, by decomposing the noise into baseband, in-phase and quadrature components as
in \eqref{eq:sec_rc_nlg_von_bbiq} and neglecting the noise terms around the second
harmonics, three linear and homogeneous differential equations for these noise terms are
obtained. The details of derivation for each term of \eqref{eq:sec_rc_nlc_von_gov_aprx}
are shown in Table~\ref{eq:sec_rc_nlc_terms}. The baseband noise equation is
\begin{flalign}\label{eq:sec_rc_nlc_vnbb_gov}
    v_{n,\textbf{BB}}(t) + RC_0\dot{v}_{n,\textbf{BB}}(t) = v_{i,n}(t) 
    - \frac12 RC_1V_{o,1}\dot{v}_{n,\textbf{I}}(t).
\end{flalign}
Unlike the nonlinear conductance case discussed before, we don't have any noise folding
from the first harmonic proportional to the noise component itself, but we see the
derivative of the quadrature component of the noise is folded back to baseband noise. 
For the in-phase component we have
\begin{flalign}\label{eq:sec_rc_nlc_vni_gov}
    v_{n,\textbf{I}}(t) + RC_0\dot{v}_{n,\textbf{I}}(t) = -RC_0\omega_0v_{n,\textbf{Q}}(t)
    - RC_1V_{o,1}\dot{v}_{n,\textbf{BB}}(t)
\end{flalign}
The in-phase noise has two source terms: The derivative of the baseband noise and the
quadrature noise. Neglecting the derivative of baseband noise and assuming the linear time
constants of RF circuit does not influence modulated noise terms, the in-phase noise is
a secondary effect and is caused by generation of quadrature noise in the nonlinear
element in the first place.

Finally, for the quadrature noise we have 
\begin{flalign}\label{eq:sec_rc_nlc_vnq_gov}
    \nonumber
    v_{n,\textbf{Q}}(t) + RC_0\dot{v}_{n,\textbf{Q}}(t) ={}&
    RC_0\omega_0v_{n,\textbf{I}}(t)  \\
    &+ RC_1\omega_0V_{o,1}v_{n,\textbf{BB}}(t).
\end{flalign}
The quadrature noise has two source terms: the in-phase noise that is
quadrature-transformed at the linear part of the capacitor and the baseband noise that is
quadrature-modulated by the nonlinear part of the capacitor. 

\begin{figure}[htb]
     \centering
\subfloat[\label{fig:plot_nlc_ham}]{
    \ifusetikzpdf
    \includegraphics[]{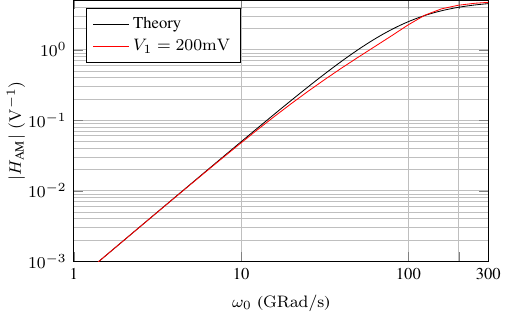}
    \else
    \tikzsetnextfilename{plot_nlc_ham}
    \input{./tikz/plot_nlc_ham.tex}
    \fi
}\\
\subfloat[\label{fig:plot_nlc_hpm}]{
    \ifusetikzpdf
    \includegraphics[]{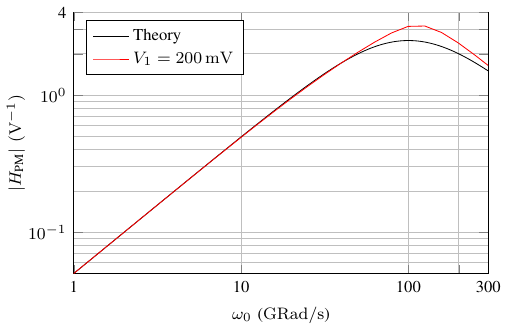}
    \else
    \tikzsetnextfilename{plot_nlc_hpm}
    \input{./tikz/plot_nlc_hpm.tex}
    \fi
}
    \caption
    {
    AM and PM noise transfer functions in RC circuit with nonlinear transcondtance;
     $R=100~\si{\Omega}$, $C_0=100~\si{fF}$, $C_1=500~\si{fF/V}$, 
 }
\label{fig:plot_nlc}
\end{figure}

By neglecting all the time derivatives of the baseband and modulated noise terms, these
equations have a simple solution
\begin{subequations}
\label{eq:sec_rc_nlc_vn} 
\begin{align}
    &v_{n,\textbf{BB}}(t) \approx v_{i,n}(t),
\label{eq:sec_rc_nlc_vnbb}\\
    &v_{n,\textbf{I}}(t) \approx 
    \frac{-R^2C_0C_1\omega_0^2} {1+(RC_0\omega_0)^2} V_{o,1}v_{i,n}(t),
\label{eq:sec_rc_nlc_vni}\\
    &v_{n,\textbf{Q}}(t) \approx 
    \frac{RC_1\omega_0} {1+(RC_0\omega_0)^2} V_{o,1}v_{i,n}(t).
\label{eq:sec_rc_nlc_vnq}
\end{align}
\end{subequations}
Therefore, the AM and PM noise transfer functions can be written as 
\begin{flalign}\label{eq:harm_nf_pn}
    H_\text{AM} = \frac{-R^2C_0C_1\omega_0^2} {1+(RC_0\omega_0)^2}.
    \quad \text{and} \quad
    H_\text{PM} = \frac{-RC_1\omega_0} {1+(RC_0\omega_0)^2}
\end{flalign}
Fig.~\ref{fig:plot_nlc} shows the simulation results of an RC circuit with nonlinear
capacitance in comparison with theory. The simulation results show that the AM/PM noise
transfer functions can be estimated with a good precision at moderate excitation levels.

\begin{figure}[htb]
    \centering
    \ifusetikzpdf
    \includegraphics[]{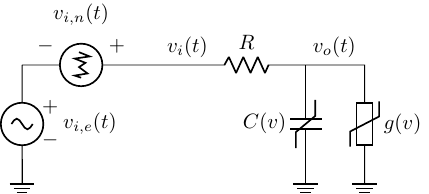}
    \else
    \tikzsetnextfilename{sch_rc_nlgc}
    \input{./tikz/sch_rc_nlgc.tex}
    \fi
    \caption{
       Nonlinear RC circuit with a nonlinear capacitance and a nonlinear conductance.
}
    \label{fig:sch_rc_nlgc}
\end{figure}

\subsection{Noise in RC circuit with nonlinear conductance and nonlinear capacitance}
After previous discussions, we are now ready to discuss the simultaneous effect of
nonlinear conductance and capacitance of a nonlinear RC circuit on the output noise 
of the circuit, shown in Fig.~\ref{fig:sch_rc_nlgc}. Although the
problem seems complicated, most of the steps to find the noise response have already been
performed. This will be done in a step by step approach similar to previous sections. It
is also a good practice to review what needs to be done for more sophisticated nonlinear
circuits.

$\bullet$ Deriving the governing equation of the circuit
\begin{flalign}\label{eq:sec_rc_nlgc_gov}
    v_{o}(t) = v_{i}(t) - RC_0 \dot{v}_{o}(t) 
    - g_2Rv_o^2(t) - RC_1v_{o}(t) \dot{v}_{o}(t).
\end{flalign}

$\bullet$ Finding the linear response of the circuit. The fundamental tone response at
moderate input levels once again is similar to linear response given in
\eqref{eq:sec_rc_nlg_vo1} and \eqref{eq:sec_rc_nlg_vo1_mag_phi}.

$\bullet$ Deriving the noise nonlinear differential equation
\begin{flalign}\label{eq:sec_rc_nlgc_von_gov}
    \nonumber
    v_{o,n}(t) ={}& v_{i,n}(t) - RC_0 \dot{v}_{o,n}(t) 
    - 2g_2Rv_{o,e}(t)v_{o,n}(t) \\
    \nonumber
    & - RC_1v_{o,e}(t) \dot{v}_{o,n}(t) 
    - RC_1\dot{v}_{o,e}(t) v_{o,n}(t) \\
    & - g_2Rv_{o,n}^2(t)
    - RC_1v_{o,n}(t) \dot{v}_{o,n}(t).
\end{flalign}

$\bullet$ Neglecting the noise self modulation terms
\begin{flalign}\label{eq:sec_rc_nlgc_von_gov_aprx}
    \nonumber
    v_{o,n}(t) ={}& v_{i,n}(t) - RC_0 \dot{v}_{o,n}(t) 
    - 2g_2Rv_{o,e}(t)v_{o,n}(t) \\
    & - RC_1v_{o,e}(t) \dot{v}_{o,n}(t) 
    - RC_1\dot{v}_{o,e}(t) v_{o,n}(t) .
\end{flalign}

$\bullet$ Decompose the noise response to baseband, in-phase and quadrature noise around
the carrier given in \eqref{eq:sec_rc_nlg_von_bbiq} and neglecting the noise terms
modulated around higher order harmonics.

$\bullet$ Substituting the assumed noise response given in \eqref{eq:sec_rc_nlg_von_bbiq}
into the noise differential equation \eqref{eq:sec_rc_nlgc_von_gov_aprx} and neglecting
the noise terms modulated around the second harmonic.

$\bullet$ Equating the based noise, in-phase- and quadrature-modulated noise terms on both
sides of the equation and obtaining a new set of linear and homogeneous differential equations
\begin{subequations}
\label{eq:sec_rc_nlgc_vn} 
\begin{align}
    \nonumber
    v_{n,\textbf{BB}}(t) + RC_0\dot{v}_{n,\textbf{BB}}(t) 
     ={}& v_{i,n}(t) - g_2RV_{o,1}v_{n,\textbf{I}}(t) \\
    &- \frac12 RC_1V_{o,1}\dot{v}_{n,\textbf{I}}(t).
\label{eq:sec_rc_nlgc_vnbb_gov}
\end{align}
\begin{align}
    \nonumber
    v_{n,\textbf{I}}(t) + RC_0\dot{v}_{n,\textbf{I}}(t) ={}& 
    -RC_0\omega_0v_{n,\textbf{Q}}(t) 
    - 2g_2RV_{o,1}v_{n,\textbf{BB}}(t) \\
    &- RC_1V_{o,1}\dot{v}_{n,\textbf{BB}}(t)
\label{eq:sec_rc_nlgc_vni_gov}
\end{align}
\begin{align}
    \nonumber
    v_{n,\textbf{Q}}(t) + RC_0\dot{v}_{n,\textbf{Q}}(t) ={}&
    RC_1\omega_0V_{o,1}v_{n,\textbf{BB}}(t) \\ 
    & RC_0\omega_0v_{n,\textbf{I}}(t) .
\label{eq:sec_rc_nlgc_vnq_gov}
\end{align}
\end{subequations}

\begin{figure*}[htb]
    \centering
\subfloat[\label{fig:sch_bipolar_setup}]{
    \ifusetikzpdf
    \includegraphics[]{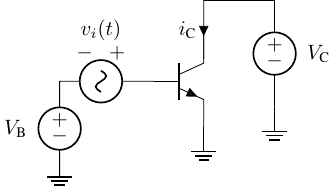}
    \else
    \tikzsetnextfilename{sch_bipolar_setup}
    \input{./tikz/sch_bipolar_setup.tex}
    \fi
}%
\subfloat[\label{fig:sch_bipolar_vbic}]{
    \ifusetikzpdf
    \includegraphics[]{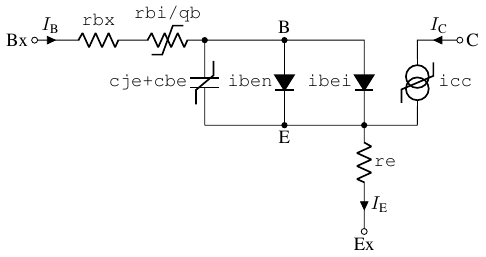}
    \else
    \tikzsetnextfilename{sch_bipolar_vbic}
    \input{./tikz/sch_bipolar_vbic.tex}
    \fi
}\\
\subfloat[\label{fig:sch_bipolar_curve_tracer}]{
    \ifusetikzpdf
    \includegraphics[]{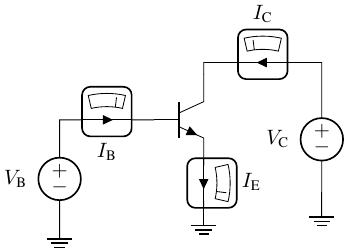}
    \else
    \tikzsetnextfilename{sch_bipolar_curve_tracer}
    \input{./tikz/sch_bipolar_curve_tracer.tex}
    \fi
}%
\subfloat[\label{fig:sch_bipolar_dc_op}]{
    \ifusetikzpdf
    \includegraphics[]{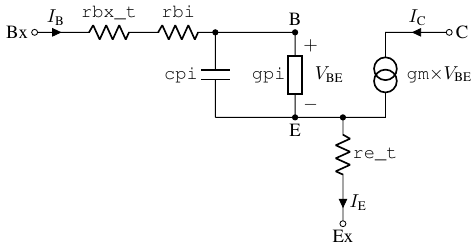}
    \else
    \tikzsetnextfilename{sch_bipolar_dc_op}
    \input{./tikz/sch_bipolar_dc_op.tex}
    \fi
}
    \caption{
        (a) Test setup for simulation of AM/PM noise in bipolar transistor, 
        (b) nonlinear model of bipolar transistor,
        (c) curve tracer setup for extraction of bipolar transistor parameters and 
        (d) small signal model of bipolar transistor. 
}
    \label{fig:sch_bipolar_model}
\end{figure*}

$\bullet$ Neglecting the effect of short time constants of the RF circuit on low-frequency
noise terms. This means all time derivatives of the noise terms can be neglected. Also
neglecting the in-phase modulated noise that is folded back to baseband frequencies 
(the term $g_2RV_{o,1}v_{n,\textbf{I}}(t)$ in \eqref{eq:sec_rc_nlgc_vni_gov}). The solution
of this set of equation will be
\begin{subequations}
\begin{align}
    &v_{n,\textbf{BB}}(t) = v_{i,n}(t)
\label{eq:sec_rc_nlgc_vnbb}\\
    &v_{n,\textbf{I}}(t) = 
    -\frac{2g_2R + R^2C_0C_1\omega_0^2} {1+(RC_0\omega_0)^2}V_{o,1}v_{i,n}(t)
\label{eq:sec_rc_nlgc_vni}\\
    &v_{n,\textbf{Q}}(t) =
    \frac{ RC_1\omega_0 - 2g_2R^2C_0\omega_0} {1+(RC_0\omega_0)^2}V_{o,1}v_{i,n}(t)
\label{eq:sec_rc_nlgc_vnq}
\end{align}
\end{subequations}
$\bullet$ Extracting the AM and PM noise transfer function from the solution
\begin{subequations}
\label{eq:sec_rc_nlgc_hamhpm} 
\begin{align}
    H_\text{AM} &= -\frac{2g_2R + R^2C_0C_1\omega_0^2} {1+(RC_0\omega_0)^2}
\label{eq:sec_rc_nlgc_ham} \\
    H_\text{PM} &= \frac{ -RC_1\omega_0 + 2g_2R^2C_0\omega_0} {1+(RC_0\omega_0)^2} 
\label{eq:sec_rc_nlgc_hpm}
\end{align}
\end{subequations}
Equation \eqref{eq:sec_rc_nlgc_hamhpm} shows that the additive phase noise of nonlinear RC
circuit depends on the sign of 2nd order nonlinear coefficients $g_2$ and $C_1$. This
behavior can potentially be used to reduce the overall PM noise of active circuits.

\section{Applying the theory to bipolar transistor}
\label{sec:bipolar}
Now that we discussed the AM and PM noise in a relatively simple nonlinear circuit, we
are ready to apply our
theory to a more practical case. We choose a common-emitter bipolar transistor for this
purpose, shown in Fig.~\subref*{fig:sch_bipolar_setup}. The problem description seems simple:
what is the additive AM and PM noise at the output collector current? However, without the
mathematical calculations we did in the previous sections, it would have been a very complex
problem.

For our simulation, we use a SiGe bipolar transistor offered from \si{130~nm} technology
node from IHP. 
The transistor is modeled according to Vertical Bipolar Intercompany Model (VBIC). In
order to simplify the complex model parameters, we neglect the effect of parasitic
pnp transistor as well as external base-emitter junction. We also neglect the effect of
feedback base-collector capacitance and Early effect. The simplified model of transistor
is illustrated in Fig.~\subref*{fig:sch_bipolar_vbic}. In order to further simplify our
analysis, we employ a simulation-assisted parameter extraction with Virtuoso analog design
environment commercial software from Cadence Design Systems, Inc. 
In the next section, we explain how the second order nonlinearity coefficients are extracted
with a curve tracer setup. For an enthusiastic reader that is interested in regeneration
of the results, we used two different notations. For instance, when we refer to our design
parameters such as the base-collector transconductance, we use $g_m$. When we refer to a
standard parameter that is provided by the model or the simulation software, we use the
same exact notation used in the software, such as \texttt{gm} for base-collector
transconductance. The relation between our model parameters and the simulation software
parameters are given via equations.

\begin{figure*}[htb]
     \centering
\subfloat[\label{fig:plot_bipolar_gpi1gpi2}]{
    \ifusetikzpdf
    \includegraphics[]{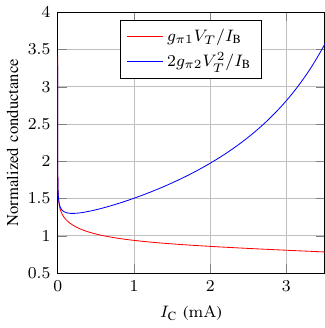}
    \else
    \tikzsetnextfilename{plot_bipolar_gpi1gpi2}
    \input{./tikz/plot_bipolar_gpi1gpi2.tex}
    \fi
}%
\subfloat[\label{fig:plot_bipolar_gm1gm2}]{
    \ifusetikzpdf
    \includegraphics[]{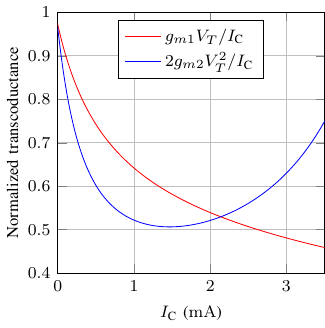}
    \else
    \tikzsetnextfilename{plot_bipolar_gm1gm2}
    \input{./tikz/plot_bipolar_gm1gm2.tex}
    \fi
}%
\subfloat[\label{fig:plot_bipolar_cpi1cpi2}]{
    \ifusetikzpdf
    \includegraphics[]{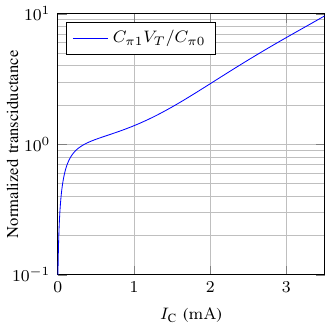}
    \else
    \tikzsetnextfilename{plot_bipolar_cpi1cpi2}
    \input{./tikz/plot_bipolar_cpi1cpi2.tex}
    \fi
}
    \caption
    {
        Normalized linear and 2nd order nonlinear coefficents extraxted from curve tracer
        setup; (a) base-emitter conductance, (b) base-collector transconductance, and (c)
        base-emitter capacitance.
 }
\label{fig:param}
\end{figure*}

\subsection{Curve tracer for large signal parameter extraction}
In order to extract the 2nd order nonlinear coefficients necessary for our analysis, we
perform
a simulation-assisted parameter extraction, rather than trying to solve the nonlinear
equations of bipolar transistor at the desired operating point. This approach
significantly simplifies the raw material required for our nonlinear analysis. We set up a
curve tracer, shown in Fig.~\subref*{fig:sch_bipolar_curve_tracer} and use the software-extracted
small signal parameters shown in Fig.~\subref*{fig:sch_bipolar_dc_op}.
For
instance, instead of finding \texttt{cje} and \texttt{cbe} at the desired operating point,
we directly use the small signal capacitance value \texttt{cpi} provided by the simulation
software. The value of this capacitance is a function of the device operating point. The
derivative of this capacitance with respect to base-emitter voltage then gives the first
nonlinear coefficient of the capacitance. Therefore, By
sweeping the base-emitter voltage, this nonlinear coefficient can be extracted. However,
we need to perform the derivative operation with respect to base-emitter voltage of the
\emph{inner} transistor, not the voltage applied to the external junctions of the base and
the emitter. Therefore, the effect of base and emitter parasitic should be
de-embedded. The base resistor is composed of two terms, the internal
base resistor \texttt{rbi} which is the linearized version of the nonlinear
\texttt{rbi/qb} resistor, and external ohmic and linear resistor
\texttt{rbx\_t}. For the emitter resistor we have just an ohmic resistor \texttt{re\_t}.
The base and emitter resistances therefore are
\begin{flalign}\label{eq:ct_rbre}
    R_\text{B} = \texttt{rbx\_t} + \texttt{rbi} \quad \text{and} \quad
    R_\text{E} = \texttt{re\_t} .
\end{flalign}
The inner base-emitter voltage can consequently be written as
\begin{flalign}\label{eq:ct_vbe}
    V_\text{BE} = V_\text{BxEx} - R_\text{B} I_\text{B} - R_\text{E} I_\text{E}
\end{flalign}

For extraction of nonlinear coefficient of base conductance $g_{\pi}(V_\text{BE})$ we
start from the first derivative of base current (with respect to $V_\text{BE}$) provided
by dc analysis, \texttt{gpi}.
\begin{flalign}\label{eq:ct_gpi1gpi2}
    \begin{dcases}
        g_{\pi1} =  \texttt{gpi}\\
        g_{\pi2} = \frac{1}{2}\frac{\text{d}\texttt{gpi}}{\text{d}V_\text{BE}}
    \end{dcases}
\end{flalign}
The normalized conductance coefficients, $g_{\pi1}$ and $g_{\pi2}$, are plotted in
Fig.~\subref*{fig:plot_bipolar_gpi1gpi2}. Similarly, the linear and 2nd order nonlinear
coefficients of the base-collector
transconductance and base-emitter capacitance can be found as
\begin{flalign}\label{eq:ct_gm1gm2}
    \begin{dcases}
        g_{m1} =  \texttt{gm}\\
        g_{m2} = \frac{1}{2}\frac{\text{d}\texttt{gm}}{\text{d}V_\text{BE}}
    \end{dcases}
\end{flalign}

\begin{flalign}\label{eq:ct_cpi0cpi1}
    \begin{dcases}
        C_{\pi0} = \texttt{cpi} \\
        C_{\pi1} = 2\frac{\text{d}\texttt{cpi}}{\text{d}V_\text{BE}}
    \end{dcases}
\end{flalign}
The normalized transconductance and capacitance coefficients are plotted in
Fig.~\subref*{fig:plot_bipolar_gm1gm2} and Fig.~\subref*{fig:plot_bipolar_cpi1cpi2} for a
transistor in IHP \si{130~\nano\meter} technology with an emitter length of
\si{0.48~\micro\meter}.

\begin{figure*}[t]
    \centering
\subfloat[\label{fig:sch_bipolar_largesignal}]{
    \ifusetikzpdf
    \includegraphics[]{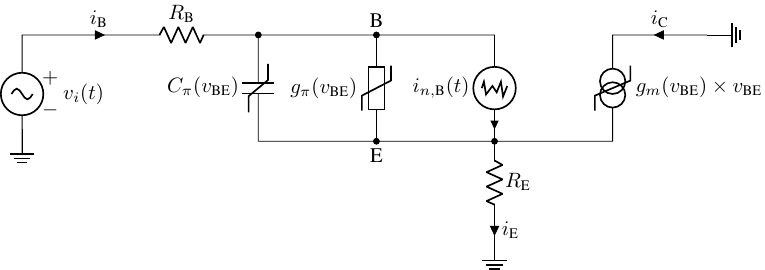}
    \else
    \tikzsetnextfilename{sch_bipolar_largesignal}
    \input{./tikz/sch_bipolar_largesignal.tex}
    \fi
}\\ 
\subfloat[\label{fig:sch_bipolar_eq_nlgc}]{
    \ifusetikzpdf
    \includegraphics[]{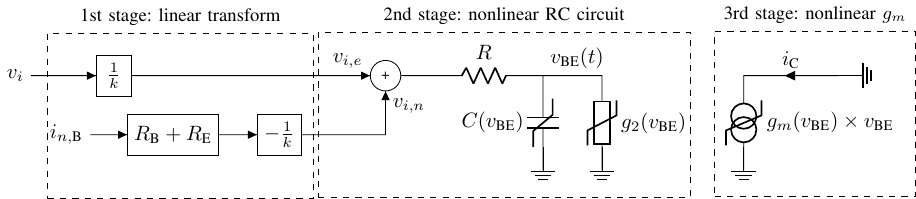}
    \else
    \tikzsetnextfilename{sch_bipolar_eq_nlgc}
    \input{./tikz/sch_bipolar_eq_nlgc.tex}
    \fi
}\\
\subfloat[\label{fig:plot_bipolar_hpm}]{
    \ifusetikzpdf
    \includegraphics[]{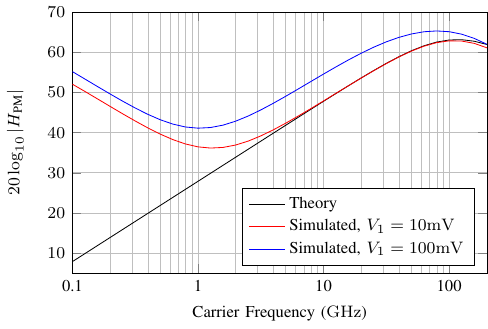}
    \else
    \tikzsetnextfilename{plot_bipolar_hpm}
    \input{./tikz/plot_bipolar_hpm.tex}
    \fi
}
\subfloat[\label{fig:plot_bipolar_ham}]{
    \ifusetikzpdf
    \includegraphics[]{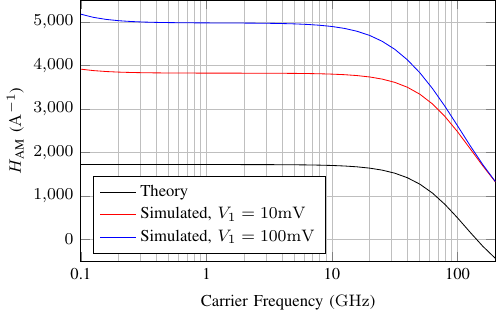}
    \else
    \tikzsetnextfilename{plot_bipolar_ham}
    \input{./tikz/plot_bipolar_ham.tex}
    \fi
}
    \caption{
        (a) Simplified nonlinear model of bipolar transistor, (b) its equivalent
        circuit for our nonlinear analysis, (c) PM noise transfer function and (d) AM
        noise transfer function. 
}%
    \label{fig:sch_bipolar_nlgc}%
\end{figure*}

\subsection{AM and PM noise transfer functions of bipolar transistor}
Figure~\subref*{fig:sch_bipolar_largesignal} shows the large-signal model of our simple
test setup, which was presented in Fig.~\subref*{fig:sch_bipolar_setup}, including its 2nd
order nonlinear
coefficients and the flicker noise source of base-emitter junction $i_{n,\text{B}}$.
The transistor is biased at \si{1~\milli\ampere} collector current.
The governing equation for $v_\text{BE}$ can be found using the following set of equations
\begin{subequations}
\begin{align}
    &v_\text{BE}(t) = v_i(t) - R_\text{B}i_\text{B}(t) - R_\text{E}i_\text{E}(t) 
\label{eq:bipolar_subgov_vbe} \\
    &i_\text{C}(t) = g_m\big(v_\text{BE}(t)\big) \times v_\text{BE}(t) 
\label{eq:bipolar_subgov_ic} \\
    &i_\text{E}(t) = i_\text{B}(t) + i_\text{C}(t) 
\label{eq:bipolar_subgov_ie} \\
    &i_\text{B}(t) = g_{\pi}\big(v_\text{BE}(t)\big) \times v_\text{BE}(t)
    + C_{\pi}\big(v_\text{BE}(t)\big)\dot{v}_\text{BE}(t) + i_{n,\text{B}}(t) 
\label{eq:bipolar_subgov_ib} 
\end{align}
\end{subequations}
The dependence of the transistor parameters to time are clear in the equations above. From
this point on, for notation simplicity, we avoid writing this time-dependence and, for
instance, write $v_\text{BE}$ instead of $v_\text{BE}(t)$ or 
$g_m$ instead of $g_m\big(v_\text{BE}(t)\big)$. First we substitute the collector current from
\eqref{eq:bipolar_subgov_ic} to the right-hand side emitter current equation in
\eqref{eq:bipolar_subgov_ie} and eliminate
$i_\text{C}$ From our equations. Then we substitute the base current and emitter currents
in \eqref{eq:bipolar_subgov_ie} and \eqref{eq:bipolar_subgov_ib} to the base-emitter voltage
equation in \eqref{eq:bipolar_subgov_vbe}. This may sound trivial, but we emphasize on this
step-by-step approach to remind that our system is nonlinear and most the equations from
linear system analysis are not applicable here.  The substitution explained above results
in the differential equation of $v_\text{BE}$ as
\begin{flalign}\label{eq:bipolar_gov_general}
    \nonumber
    v_\text{BE} ={}& v_i - (R_\text{B} + R_\text{E})i_\text{n,B} 
    - g_{\pi}(R_\text{B} + R_\text{E})v_\text{BE} \\
    & - g_{m}R_\text{E}v_\text{BE}
    - (R_\text{B} + R_\text{E})C_{\pi}\dot{v}_\text{BE}
\end{flalign}
Depending on our degree of approximation for $g_{m}$, $g_{\pi}$ and $C_{\pi}$, this equation
could be linear or nonlinear. Now we use the 2nd order nonlinear coefficient extracted by
the curve tracer setup in the previous section
\begin{subequations}
\begin{align}
    &g_m(v_\text{BE}) = g_{m1} + g_{m2}v_\text{BE} ,
\label{eq:bipolar_gov_gm} \\
    &g_{\pi}(v_\text{BE}) = g_{\pi1} + g_{\pi2}v_\text{BE} ,
\label{eq:bipolar_gov_gpi} \\
    &C_{\pi}(v_\text{BE}) = C_{\pi0} + C_{\pi1}v_\text{BE} .
\label{eq:bipolar_gov_cpi} 
\end{align}
\end{subequations}
Substituting these nonlinear coefficients into \eqref{eq:bipolar_gov_general} leads to 
nonlinear differential equation of 
\begin{flalign}\label{eq:bipolar_gov_vbe}
    \nonumber
    (1+&g_{\pi1}(R_\text{B} + R_\text{E})+g_{m1}R_\text{E})v_\text{BE} = v_i \\
    \nonumber
    &- (R_\text{B} + R_\text{E})i_\text{n,B} 
    - (R_\text{B} + R_\text{E})C_{\pi0}\dot{v}_\text{BE} \\
    \nonumber
    &- (g_{\pi2}(R_\text{B} + R_\text{E})+g_{m2}R_\text{E})v^2_\text{BE} \\
    &- (R_\text{B} + R_\text{E})C_{\pi1}v_\text{BE}\dot{v}_\text{BE}
\end{flalign}
This equation is similar to the governing equation of the RC circuit with nonlinear
conductance and nonlinear capacitance in \eqref{eq:sec_rc_nlgc_gov}. On the left-hand
side of both equations, we have just the term of unknown variable. On the right-hand side, we
have the input signal term, input noise terms, linear derivative of the output voltage
caused by the linear part of the capacitance, a nonlinear term caused by the overall
nonlinear conductance and transconductance terms in transistor, and, a nonlinear term
proportional to the output voltage and its derivative cause by the nonlinear part of the
capacitance. In order to make these equations exactly identical, we define the following
set of parameters.
\begin{subequations}
\allowdisplaybreaks%
\label{eq:bipolar_params} 
\begin{align}
    &v_{i,e} := \frac1k v_i 
\label{eq:bipolar_param_vie} \\
    &v_{i,n} := -\frac1k (R_\text{B} + R_\text{E})i_\text{n,B} 
\label{eq:bipolar_param_vin} \\
    &k := 1+g_{\pi1}(R_\text{B} + R_\text{E})+g_{m1}R_\text{E}
\label{eq:bipolar_param_k} \\
    &C_0 := C_{\pi0} 
\label{eq:bipolar_param_c0} \\
    &C_1 := C_{\pi1} 
\label{eq:bipolar_param_c1} \\
    &R := \frac1k (R_\text{B} + R_\text{E}) 
\label{eq:bipolar_param_r} \\
    &g_2 := \frac{g_{\pi2}(R_\text{B} + R_\text{E})+g_{m2}R_\text{E}}
    {R_\text{B} + R_\text{E}}
\label{eq:bipolar_param_g2}
\end{align}
\end{subequations}
With this set of parameters, we can find the AM and PM noise transfer functions for the
voltage across the inner base-emitter diode terminals. But we want to find the AM and PM
noise transfer functions with respect to the collector current. Besides, the input signal
signal is scaled by $1/k$ and the base-emitter noise current is scaled with $(R_\text{B} +
R_\text{E})/k$. So we have to take into account the effect of linear scaling of the
inputs. These linear and nonlinear transformation are shown in
Fig.~\subref*{fig:sch_bipolar_eq_nlgc} as different stages. Now we find the overall AM and
PM noise transfer functions. We start with $H_\text{PM}$. The last stage is a memoryless
element, so it does not have any effect on the phase noise. Therefore, we can write the
output PM noise at the $v_\text{BE}$ node as 
\begin{flalign}\label{eq:bipolar_vnpm_be1}
    v_{n,\textbf{Q}, \text{BE}}(t) = - H_\text{PM,2} \times v_{\text{BE}}(t) \times v_{i,n}(t).
\end{flalign}
Substituting $v_{i,n}$ from \eqref{eq:bipolar_param_vin} in \eqref{eq:bipolar_vnpm_be1}, gives
\begin{flalign}\label{eq:bipolar_vnpm_be2}
    v_{n,\textbf{Q}, \text{BE}}(t) = 
    -H_\text{PM,2} \times v_{\text{BE}}(t)
    \times \left[ -\frac1k (R_\text{B} + R_\text{E})i_\text{n,B} \right].
\end{flalign}
Therefore, the overall PM noise transfer function is
\begin{flalign}\label{eq:bipolar_hpm}
    H_\text{PM} = 
    -\frac1k (R_\text{B} + R_\text{E})H_\text{PM,2} .
\end{flalign}
where $H_\text{PM,2}$ can be calculated using \eqref{eq:sec_rc_nlgc_hpm} and
\eqref{eq:bipolar_param_vie} to \eqref{eq:bipolar_param_g2}.
Finding the AM noise requires more caution, as the 3rd stage has 2nd order nonlinearity.
The AM noise transfer function of the third stage can be written as
\begin{flalign}\label{eq:bipolar_ham3}
    H_\text{AM,3} = 2\frac{g_{m,2}}{g_{m,1}}
\end{flalign}
The second stage has a linear gain of $v_\text{BE}/v_{i,e}=1$, as $g_2$ is purely
nonlinear of 2nd order. Therefore, the overall AM noise
transfer function of the 2nd and 3rd cascaded stages is (see appendix~\ref{apdx:cascade})
\begin{flalign}\label{eq:bipolar_ham23}
    H_\text{AM,2-3} = H_\text{AM,2} + H_\text{AM,3} 
\end{flalign}
Now we can write the AM noise of the collector current as
\begin{flalign}\label{eq:bipolar_inam_c}
    i_{n,\textbf{I}, \text{C}}(t) = 
    H_\text{AM,2-3} \times i_{\text{C}}(t)
    \times \left[ -\frac1k (R_\text{B} + R_\text{E})i_\text{n,B} \right].
\end{flalign}
Therefore we have
\begin{flalign}\label{eq:bipolar_ham}
    H_\text{AM} = 
    -\frac1k (R_\text{B} + R_\text{E})
    \left( H_\text{AM,2} + 2\frac{g_{m,2}}{g_{m,1}} \right).
\end{flalign}
where $H_\text{AM,2}$ can be calculated using \eqref{eq:sec_rc_nlgc_ham} and
\eqref{eq:bipolar_param_vie} to \eqref{eq:bipolar_param_g2}. We avoid further
approximation and compare the theoretical results with the simulation results, 
plotted in Figs.~\subref*{fig:plot_bipolar_hpm} and \subref*{fig:plot_bipolar_ham}. At
moderate excitation levels, $H_\text{PM}$ shows a good matching with theory at high
frequencies. The difference between the theoretical and simulation results at low
frequencies is due to the low-frequency pole of HBT transistor caused by self heating,
which was not included in our analysis. At higher excitation levels, the higher order
nonlinear coefficients affect $H_\text{PM}$ which leads to deviation from our 2nd order
nonlinear analysis. The simulated AM noise transfer function is higher than the
estimated value by almost 2-3 times at different excitation levels. This difference can be
attributed mainly to approximations related to memoryless elements; for instance,
neglecting 2nd order nonlinearity of \texttt{rbi}, the nonlinear collector-emitter
conductance caused by the Early effect, and the effect of collector parasitic pnp
transistor. Although all these effects were neglected, the analytical approach still
provides a good estimation of AM and PM noise transfer functions.

\section{Conclusions} \label{sec:conclusion}
In this paper, we provided a systematic approach for understanding the nonlinear
mechanisms that lead to additive AM/PM noise generation in active devices. A mathematical
model for including the AM and PM noise in active devices was provided. Based on this
model, various noisy nonlinear circuits were analyzed and the AM/PM noise transfer
functions were extracted and the analytical results were compared with theory. Finally,
this approach was applied to a bipolar transistor, and the analytical results were
validated by comparing them with the simulation results.

\appendices

\section{Cascaded stages with second order nonlinearity}\label{apdx:cascade}
\begin{figure}[htb]
    \centering
    \ifusetikzpdf
    \includegraphics[]{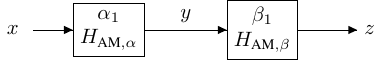}
    \else
    \tikzsetnextfilename{sch_nl_cascade}
    \input{./tikz/sch_nl_cascade.tex}
    \fi
    \caption{
        Cascaded stages with 2nd order nonlinearity.
}
    \label{fig:sch_nl_cascade}
\end{figure}

An important scenario in nonlinear noise mechanisms is having two nonlinear elements,
$\alpha$ and $\beta$, acting in series, shown in Fig.~\ref{fig:sch_nl_cascade}.

The first \& the second element have a
linear gain of $\alpha_1$ \& $\beta_1$ and AM noise transfer function of $H_{\text{AM},\alpha}$
\& $H_{\text{AM},\beta}$, respectively. We want to find the AM noise transfer function of
the overall
system. Let's assume the input has a monotone excitation part, given in \eqref{eq:basics_xie},
and a noise part as
\begin{flalign}\label{eq:nl_dev_cascade_x}
    x(t) = x_{i,e}(t) + x_{i,n}(t).
\end{flalign}
The output of the first element can be written as 
\begin{flalign}\label{eq:nl_dev_cascade_y}
    y(t) = \alpha_1 x_{i,e}(t) + \alpha_1 x_{i,n}(t) 
    + H_{\text{AM},\alpha}x_{i,n}(t) \alpha_1 x_{i,e}(t),
\end{flalign}
The third term in the right-hand side if is the AM noise modulated around the carrier. In
the third term in the right-hand side of \eqref{eq:nl_dev_cascade_y}, we
placed $\alpha_1$ intentionally behind $x_{i,e}$ to emphasize that $H_\text{AM}$ is
defined with respect to the \emph{output} of the nonlinear element. Now we apply $y(t)$ to
the
element $\beta$. All the inputs of element $\beta$ in are linearly transferred to output by
coefficient $\beta_1$ and an additional AM noise term in generated that is proportional to
$H_{\text{AM},\beta}$, the \emph{input} noise, and the \emph{output} signal of element $\beta$.
Therefore
\begin{flalign}\label{eq:nl_dev_cascade_z}
    \nonumber
    z(t) ={}& \beta_1[\alpha_1 x_{i,e}(t) + \alpha_1 x_{i,n}(t) 
    + H_{\text{AM},\alpha}x_{i,n}(t) \alpha_1 x_{i,e}(t)], \\
    & + H_{\text{AM},\beta}[\alpha_1 x_{i,n}(t)] \times [ \beta_1 \alpha_1 x_{i,e}(t) ],
\end{flalign}
Therefore, the AM noise transfer function of the two elements combined can be written as
\begin{flalign}\label{eq:nl_dev_cascade_ham}
    H_{\text{AM}} = H_{\text{AM},\alpha} + \alpha_1 H_{\text{AM},\beta}
\end{flalign}

\bibliography{./bib/ref,./bib/books}
\end{document}